\documentclass[prb,aps,twocolumn,showpacs,floats]{revtex4}
\usepackage{graphicx,amsmath,epsf,amsfonts}

\newcommand{\bx}{{\bf x}}
\newcommand{\bq}{{\bf q}}
\newcommand{\bp}{{\bf p}}
\newcommand{\bn}{{\bf n}}

\newcommand{\bk}{{\bf k}}

\newcommand{\vd}{\vec \delta}

\newcommand{\pin}{p}
\newcommand{\pout}{p'}

\renewcommand{\vec}[1]{\mathbf{#1}}

\begin{document}

\title{Anderson localization from classical trajectories}

\author{Piet W.\ Brouwer} \affiliation{Laboratory of Atomic and Solid
 State Physics, Cornell University, Ithaca, NY 14853, USA}

\author{Alexander Altland}
\affiliation{Institut f\"ur theoretische Physik, Z\"ulpicher Str. 77,
 50937 K\"oln, Germany}

\date{\today}

\begin{abstract}
 We show that Anderson localization in quasi-one dimensional
 conductors with ballistic electron dynamics, such as an array of
 ballistic chaotic cavities connected via ballistic contacts, can be
 understood in terms of classical electron trajectories only. At
 large length scales, an exponential proliferation of trajectories of
 nearly identical classical action generates an abundance of
 interference terms, which eventually leads to a suppression of
 transport coefficients. We quantitatively describe this mechanism in
 two different ways: the explicit description of transition
 probabilities in terms of interfering trajectories, and an
 hierarchical integration over fluctuations in the classical phase
 space of the array cavities.
\end{abstract}


\maketitle

\section{Introduction}

The interplay of quantum phase coherence and repeated random 
scattering is at the origin of many effects in
mesoscopic physics.\cite{kn:akkermans1995,kn:imry2002}
These effects include weak localization and
universal conductance fluctuations, both of which are small but
fundamental corrections with respect to the conductance obtained from 
Drude-Boltzmann theory. They culminate in Anderson localization, 
the phenomenon that the resistance
of a one or two-dimensional electron gas grows exponentially with
system size if the system size is sufficiently
large.\cite{kn:anderson1958,kn:anderson1979b}

Originally, Anderson localization and other mesoscopic effects were
discovered in the context of disordered metals, in which electrons
scatter off impurities with a size comparable to their
wavelength. Theoretically, quantum effects in disordered metals are
described using the `disorder average' which deals with an ensemble
of macroscopically equivalent but microscopically different impurity
configurations. The presence of impurities is not essential for the
existence of quantum effects, however.  The same effects, with the
same statistical properties, have been found to appear if the electron
motion is ballistic and chaotic, the only source of scattering of
electrons being specular reflection off the sample 
boundaries.\cite{kn:stone1995,kn:nakamura2004,foot1} Besides
being of theoretical interest for understanding the quantum properties
of systems with chaotic classical dynamics,\cite{kn:haake1991} the
case of ballistic electron motion is relevant experimentally for very
clean artificially structured two-dimensional electron gases in
semiconductor heterostructures, such as quantum dots or antidot
lattices.\cite{kn:kouwenhoven1997,kn:roukes1989,kn:ennslin1990}

Unlike disordered metals, in which impurities scatter diffractively,
electrons in ballistic conductors have a well-defined classical
dynamics. Many quantum properties of ballistic conductors
with chaotic classical dynamics can be understood in terms of 
classical mechanics by making use of well-chosen semiclassical
methods.\cite{kn:haake1991,kn:nakamura2004} A
well-known example is the Gutzwiller trace formula, which relates
the density of states to properties of periodic orbits of the 
classical dynamics.\cite{kn:gutzwiller1990}
For the conductance $G$ and its quantum corrections, there is a
variant of Gutzwiller's formula, which expresses $G$ as a double sum over 
classical trajectories $\alpha$ and $\beta$ connecting the source and 
drain contacts,\cite{kn:jalabert1990,kn:stone1995}
\begin{equation}
G = \frac{1}{2 \pi \hbar}
\int dp dq\, A_{\alpha} A_{\beta} e^{i ({\cal S}_{\alpha} - {\cal
S}_{\beta})/\hbar}.
\label{eq:Gsemi}
\end{equation}
Upon entrance and exit, the trajectories $\alpha$ and $\beta$ have the
same transverse momenta $p$ and $q$, respectively, but the position at
which they enter or exit the sample may be different. Further,
$A_{\alpha,\beta}$ and ${\cal S}_{\alpha,\beta}$ are the so-called
``stability amplitude'' and the classical action of the trajectories.

With Eq.\ (\ref{eq:Gsemi}) as a starting point, the conductance, 
including its quantum corrections, can be calculated solely
from knowledge of classical trajectories. Quantum effects 
%such as weak localization or universal conductance fluctuations 
have been linked to the existence of families of classical
trajectories that only differ near `small-angle encounters' of the
trajectory with 
itself.\cite{kn:sieber2001,kn:sieber2002,kn:richter2002,kn:mueller2007,foot2}
This semiclassical
approach has been very successful explaining quantum effects in the
`perturbative regime' in which quantum interference
provides a small correction to the classical 
mechanics.\cite{kn:sieber2001,kn:sieber2002,kn:richter2002,kn:mueller2004,kn:mueller2005,kn:heusler2006,kn:whitney2006,kn:brouwer2006,kn:mueller2007}
Weak localization and universal conductance fluctuations are examples
of such perturbative quantum effects.
%Examples of quantum effects in the perturbative regime are the 
%spectral form factor $K(t)$ if $t$ is smaller than the
%Heisenberg time $\tau_{\rm H}$, or for the weak-localization
%correction to the conductance of a chaotic cavity.
%In this ``perturbative regime'', the
%semiclassical approach has reproduced results previously obtained for 
%disordered systems. In fact, the structure of the semiclassical theory
%closely parallels that of the ``diagrammatic perturbation theory''
%used to describe disordered metals. 
Recently, Heusler {\em et al.} showed that it is also possible to 
understand quantum effects in terms of classical trajectories 
outside the perturbative regime. They considered the
spectral form factor $K(t)$ of a ballistic cavity, the Fourier
transform of the two-point correlation function of the density of
states. A direct semiclassical evaluation of $K(t)$ using Gutzwiller's
trace formula was known to be
possible in the perturbative regime $t < t_{\rm H}$ only, where the
Heisenberg time $t_{\rm H} = h/\Delta$, $\Delta$ being the cavity's
mean level 
spacing.\cite{kn:berry1985,kn:aleiner1997,kn:sieber2001,kn:sieber2002,kn:mueller2004,kn:mueller2005} 
Motivated by the
field-theoretical formulation of the problem, Heusler {\em et al.}
used a different way to express $K(t)$ in terms of periodic orbits 
that allowed them
to calculate $K(t)$ for all times.\cite{kn:heusler2007}

In this communication, we consider Anderson localization for ballistic
electron gases, and show that it, too, can be understood in terms of
interference of
classical trajectories, with Eq.\ (\ref{eq:Gsemi}) as a starting
point. Examples of systems that may exhibit such `ballistic Anderson 
localization' are antidot lattices or arrays of chaotic cavities.
Although it is
generally accepted that Anderson localization exists irrespective of
the details of the microscopic electronic dynamics,\cite{foot3}
the similarity of the phenomena for ballistic and disordered
electron systems should not obscure the vastly 
different starting points of the theories for the two cases.
This difference not only pertains to the microscopic dynamics
(classical-deterministic vs.\ quantum-probabilistic), but
also to the statistical assumptions of the theory. Unlike theories
of quantum transport in disordered metals, semiclassical theories of
ballistic conductors are
intended to describe one specific system.\cite{kn:haake1991} 
Fluctuations appear solely from
variations of the Fermi energy; no changes in the classical
dynamics are invoked.

%Anderson localization not only appears in the context of electronic
%transport; It also occurs for random scattering of light or
%microwaves \cite{kn:john1984}. Dynamic localization, 
%Anderson localization in 
%momentum space rather than in real space, is relevant for 
%certain dynamic systems with a periodic
%time-dependent Hamiltonian, such as the kicked rotor
%\cite{kn:fishman1982}.

For disordered metals, Anderson localization is most prominent for a
quasi-one dimensional geometry, with sample length $L$ much larger 
than the sample
width $W$. 
For quasi-one dimensional samples, a full theory of transport in the
localized regime was developed by Dorokhov\cite{kn:dorokhov1982} and
Mello, Pereyra, and Kumar,\cite{kn:mello1988} using a stochastic
approach, and by Efetov and 
Larkin,\cite{kn:efetov1983,kn:efetov1983b} using a field-theoretic
approach. Our theory of ballistic Anderson localization closely
follows these approaches. 
The fact that the semiclassical theory for ballistic electrons follows
the corresponding quantum mechanical theory for disordered metals is
not special to the present case. It is also typical of the
trajectory-based semiclassical theories in the perturbative regime,
which have a structure that resembles the diagrammatic perturbation
theory of quantum corrections in disordered 
metals.\cite{kn:sieber2001,kn:sieber2002,kn:richter2002}

In addition to the ballistic electron gases considered here, Anderson
localization also occurs in certain dynamic systems with a periodic
time-dependent Hamiltonian, such as the kicked 
rotor.\cite{kn:haake1991} Classically, the dynamics of the kicked rotor is
chaotic, with a momentum that changes diffusively under the influence
of the periodic kicks. Quantum mechanically, it exhibits `dynamic 
localization', Anderson localization in 
momentum space rather than in real space.\cite{kn:fishman1982}
The phenomenology of dynamic localization is equal to that of
Anderson localization in disordered metals. This is confirmed by
extensive numerical
simulations,\cite{kn:casati1979,kn:shepelyanski1986}
as well as a field
theoretical analysis of the problem.\cite{kn:altland1996b}

\begin{figure}
\includegraphics[width=\hsize]{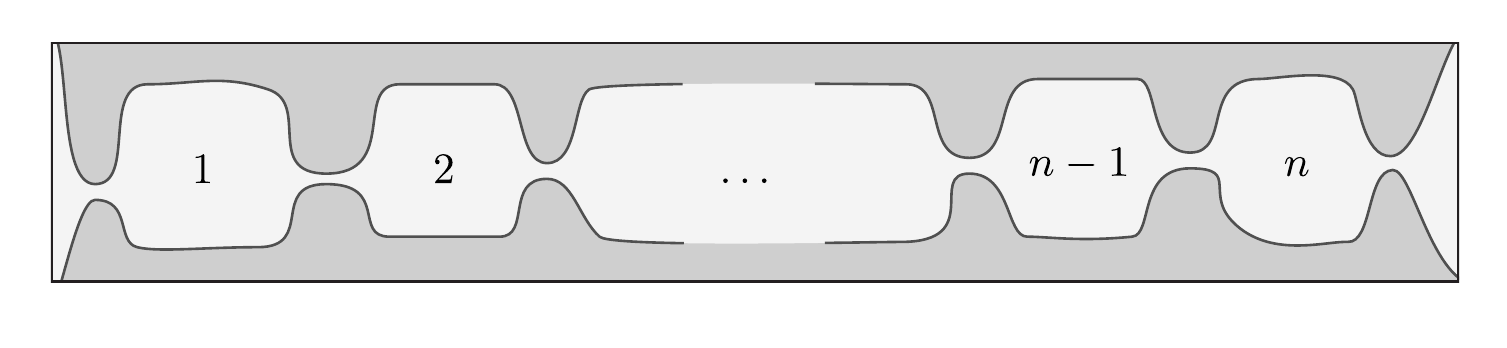}
%\epsfxsize=0.7\hsize
%\epsffile{fig1b.eps}\
\caption{\label{fig:1} (Color online)
Schematic drawing of an array of $n$ chaotic cavities. The cavities are
connected via ballistic contacts with dimensionless conductance
$g_{\rm c}$. The semiclassical theory requires the limit $g_{\rm c}
\to \infty$ at a fixed ratio $n/g_{\rm c}$.}
\end{figure}

The trajectory-based theory of ballistic Anderson localization that we
report here is constructed for a specific system, an array 
of chaotic cavities. A schematic drawing of
such an array is shown in Fig.\ \ref{fig:1}. We use
this model system because the trajectory-based theory of transport through 
a single chaotic cavity is well established in the 
literature.\cite{kn:jalabert1990,kn:baranger1993,kn:baranger1993b,kn:richter2002,kn:heusler2006,kn:mueller2007,kn:brouwer2006}
%One advantage of this
%model is that the classical dynamics inside each cavity is ergodic, so
%that we can make use of the extensive literature on the
%trajectory-based theory of transport through a single chaotic cavity 
%\cite{kn:jalabert1990,kn:baranger1993,kn:baranger1993b,kn:richter2002,kn:heusler2006,kn:mueller2007}.
Arrays of cavities have been used as a starting point for a
field-theoretic description of Anderson localization using random
matrix theory,\cite{kn:mirlin1994,kn:brouwer1996b} but 
not with Dorokhov's method.
%Anderson
%localization in an array of cavities has not been studied using
%Dorokhov's method. 
%The reason is that Dorokhov's approach relies 
%on the possibility to increase the sample length $L$ by an amount 
%$\delta L$ such that the probability of back-reflection in the added
%segment is small. The sample size of an array of cavities can only be
%increased by at least one cavity at a time. The probability of 
%back-reflection from a cavity is $1/2$, which is too large for a direct 
%application of Dorokhov's approach to an array of chaotic cavities.
%On the other hand, this probability is still small enough that 
%Dorokhov's approach can be adapted to this case [see Sec.\ \ref{sec:dis}].
%The remainder of this article is organized as follows. 
In Sec.\
\ref{sec:dis} we show how Dorokhov's theory can be adapted to this
system if the cavities are disordered and random matrix theory can be
used to describe transport through a single cavity. 
%In Sec.\ \ref{sec:dis} we also review the connection between Dorokhov's
%stochastic approach and the field-theoretic approach
%\cite{kn:brouwer1996b}.
%This theory is based on a recursion relation for
%the scattering matrix $S^{\rm q}(n)$, where $n$ is the number of
%cavities in the array. 
In Sec.\ \ref{sec:2} we summarize the basic elements of the
trajectory-based semiclassical formalism.  This formalism is used to
construct the trajectory-based semiclassical theory of ballistic
localization in Sec.\ \ref{sec:semi}. In section
\ref{sec:field_theory} we approach the localization phenomenon from a
different perspective. We describe the array in terms of a nonlinear
$\sigma$-model whose perturbative (`diagrammatic') evaluation
generates structures of paired Feynman amplitudes similar to those
appearing in the native semiclassical approach. Alternatively, the
dynamical structure of the theory can be analyzed to identify and
successively eliminate hierarchies of different types of dynamical
fluctuations in the system. In this way we obtain an effective low
energy theory which turns out to be equivalent to the nonlinear
$\sigma$-model of diffusive quantum wires; the latter model is known
to predict exponential localization at large length scales.
We conclude in Sec.\ \ref{sec:concl}.

%it is important to clarify the goal of such a
%theory. Since it is generally accepted that the transition to Anderson 
%localization is independent of the details of the microscopic electron 
%dynamics (ballistic--chaotic versus disordered--diffusive), any
%satisfactory semiclassical theory of Anderson localization is expected 
%to reproduce the existing theories in disordered metals. Thus, a 
%semiclassical theory will not explain new phenomena; It can only offer
%an alternative explanation of these phenomena, using the classical
%dynamics as its only input. More precisely, the goal of the present
%calculation is to show how the same semiclassical formalism that has 
%been able to successfully reproduce quantum effects in the 
%perturbative regime can also be used as the basis of a description of
%Anderson localization in quasi-one dimensional conductors.

\section{Array of disordered cavities}
\label{sec:dis}

We now describe how Dorokhov's approach can be used to describe
Anderson localization in an array of disordered cavities or quantum
dots.
%Before we turn to the semiclassical theory of an array of ballistic
%cavities, we first summarize the 
%theory of localization in an array of {\em disordered}
%cavities. 
We take the disorder in each cavity to be weak (cavity size
much smaller than the localization length), so that
transport through a single cavity is described by random matrix
theory.\cite{kn:beenakker1997}

A schematic drawing of the array of cavities is shown in Fig.\
\ref{fig:1}.
The cavities are connected via ballistic contacts with
dimensionless conductance $g_{\rm c}$. Since we want to compare with a
semiclassical theory for an array of ballistic cavities, we require
$g_{\rm c} \gg 1$. Localization takes place if the conductance of the
array is of order unity. This condition is met if the number $n$ of
cavities in the array is comparable to $g_{\rm c}$. Hence, in the
calculations below we take the limit $g_{\rm c} \to \infty$ while
keeping the ratio $n/g_{\rm c}$ fixed. The same limit is taken in the
field theoretical description of 
localization,\cite{kn:efetov1983,kn:efetov1983b,kn:zirnbauer1992,kn:mirlin1994,kn:brouwer1996b}
where it is known as the ``thick wire limit''.

The transport properties of the array of cavities are encoded
in its scattering matrix $S^{\rm q}(n)$. (The superscript ``q'' is
used to distinguish the quantum mechanical scattering matrix from its
semiclassical counterpart to be introduced in Sec.\ \ref{sec:2}.) 
The matrix indices of $S^{\rm q}(n)$ represent
the two contacts $i=1,2$ at the far left and right of the array
and the transverse modes in each contact,
$|p_{\perp,i}| = \pm \pi \hbar n_i/W$, $n_i=1,2,\ldots,N$, where
$N = g_{\rm c}$ is the number of channels in contact $i$, $i=1,2$. 
%The
%dimensionless conductance of the array of cavities reads
%\begin{equation}
%  g = \mbox{tr}\, S_{12}^{\rm q}(n) S_{12}^{{\rm q}\dagger}(n).
%\end{equation}
The matrix $S^{\rm q}(n)$ is a random quantity because it depends
on the Fermi energy and on the precise disorder configuration in each
cavity. Following the approach of Dorokhov, Mello, Pereyra, and 
Kumar,\cite{kn:dorokhov1982,kn:mello1988} we consider the hermitian matrix
\begin{equation}
{\cal T}^{\rm q}(n) = S_{12}^{\rm q}(n) S_{12}^{{\rm q}\dagger}(n),
\end{equation}
and calculate its statistical distribution 
by expressing ${\cal T}^{\rm q}(n)$ in terms of ${\cal T}^{\rm q}(n-1)$
and proceeding recursively. The matrix ${\cal T}^{\rm q}$ is related
to the dimensionless conductance $g(n)$ of the array through the Landauer
formula,
\begin{equation}
g(n) = \mbox{tr}\, {\cal T}^{\rm q}(n).
\end{equation}

%Here we take the scattering matrix of 
%each individual cavity from the circular unitary ensemble of random
%matrix theory. This is appropriate for a weakly disordered chaotic
%cavity. 
Taking the scattering matrix of each individual cavity from the
circular ensemble of random matrix theory,\cite{kn:beenakker1997}
one then finds that the
recursion relation for ${\cal T}^{\rm q}$ takes the form
\begin{eqnarray}
\delta {\cal T}^{\rm q} &=& {\cal T}^{\rm q}(n) - {\cal T}^{\rm q}(n-1) \nonumber \\ &=&
- \frac{1}{g_{\rm c}} {\cal T}^{\rm q}(n-1) \mbox{tr}\, {\cal
T}^{\rm q}(n-1)
- \frac{1}{g_{\rm c}} \delta_{\beta,1} {\cal T}^{\rm q}(n-1)^2
 \nonumber \\ && \mbox{}
+ {\cal X}^{\rm q}(n) + {\cal O}(g_{\rm c}^{-3/2}),
\label{eq:ttdiff}
\end{eqnarray}
where the hermitian matrix ${\cal X}$ is a random (noise) term with a Gaussian
distribution,
\begin{eqnarray}
\label{eq:calTstat} 
\langle {\cal X}^{\rm q}(n)_{ij} \rangle_n &=& 0,\\ 
\langle {\cal X}^{\rm q}(n)_{ij} {\cal X}^{\rm q}(n)_{kl} \rangle_n &=&
\frac{1}{g_{\rm c}} 
{\cal T}^{\rm q}(n-1)_{il} 
 {\cal F}^{\rm q}_{kj}
%  [{\cal T}^{\rm q}(n-1) - {\cal T}^{\rm
%  q}(n-1)^2]_{kj} 
\nonumber \\ && \mbox{} 
+
\frac{1}{g_{\rm c}} 
{\cal T}^{\rm q}(n-1)_{kj} 
 {\cal F}^{\rm Q}_{il}
%  [{\cal T}^{\rm q}(n-1) - {\cal
% T}^{\rm q}(n-1)^2]_{il}
 \nonumber \\ && \mbox{} 
+ \frac{2}{g_{\rm c}} \delta_{\beta,1}
{\cal G}^{\rm q}_{ij} {\cal G}^{{\rm q}*}_{kl},
\nonumber
\end{eqnarray}
where ${\cal F}^{\rm q} = {\cal T}^{\rm q} (1 - {\cal T}^{\rm q})$,
$\beta=1$ or $2$ in the presence or absence of time-reversal
symmetry, respectively, and
\begin{equation}
{\cal G}^{\rm q} = S_{12}^{\rm q} S_{22}^{{\rm q}\dagger}
S_{21}^{\rm q}.
\end{equation}
The averaging brackets $\langle \ldots \rangle_n$ denote an average
with respect to the disorder configuration in the $n$th cavity only.

In the limit $g_{\rm c} \to \infty$ while keeping $n/g_{\rm c}$ fixed,
the stochastic recursion relation (\ref{eq:ttdiff}) can be mapped to
the Dorokhov-Mello-Pereyra-Kumar (DMPK)
equation, which is a stochastic differential 
equation for the eigenvalues of 
${\cal T}^{\rm q}$.\cite{kn:dorokhov1982,kn:mello1988,kn:beenakker1997,foot4}
%This particular combination can be expressed in terms of ${\cal
%  T}^{\rm q}$, since
%$$
%%\begin{equation}
%  {\cal G}({\cal T}^{\rm T})^n {\cal G}^{\dagger} =
%  {\cal T}^{n+2} - {\cal T}^{n+3},\ \
%  n=0,1,2,\ldots.
%$$
%%\end{equation}
The solution of the DMPK equation is known,\cite{kn:beenakker1997}
which completes the theory of localization for an array of disordered 
cavities.

Alternatively, the stochastic recursion relation (\ref{eq:ttdiff}) 
can be used to generate a coupled set of recursion relations for 
the disorder averages of the moments of ${\cal T}(n)$,
%which read \cite{kn:tartakovski1995,kn:brouwer1998b}
\begin{widetext}
\begin{eqnarray}
\delta \left\langle \prod_{m=1}^{n} T_{i_m} \right \rangle &=&
\left\langle \prod_{m=1}^{n} T_{i_m}(n) \right \rangle 
- \left\langle \prod_{m=1}^{n} T_{i_m}(n-1) \right \rangle 
\nonumber \\ &=&
- \frac{1}{g_{\rm c}} \delta_{\beta,1} \sum_{k=1}^{n} i_{k}
\left\langle 
T_{i_{k}+1}\prod_{{m=1} \atop {m \neq k}}^{n} 
T_{i_m} \right \rangle
- \frac{1}{g_{\rm c}}
\left(\sum_{k=1}^{n} i_k \right) 
\left\langle T_1 \prod_{m=1}^{n} T_{i_m} \right\rangle
\nonumber \\ && \mbox{}
+ \frac{1}{g_{\rm c}} \sum_{k=1}^{n} \sum_{j=1}^{i_k-1} i_k 
\left\langle (T_{j} (T_{i_k-j} - T_{i_k-j+1}))
\prod_{{m=1 \atop m \neq k}}^{n} T_{i_m} \right\rangle
\nonumber \\ && \mbox{}
+ \frac{4}{\beta g_{\rm c}} 
\sum_{k=1}^{n} \sum_{l=1}^{k-1} i_{k} i_{l}
\left\langle (T_{i_{k}+i_{l}} - T_{i_{k}+i_{l}+1})
\prod_{{m=1 \atop m \neq k, l}}^{n} T_{i_m} \right\rangle
+ {\cal O}(g_{\rm c}^{-2}),
\label{eq:Tgeneral}
\end{eqnarray}
\end{widetext}
with
% $2 L/\xi = n/g_{\rm c}$ and
\begin{equation}
T_m = \mbox{tr}\, ({\cal T}^{\rm q})^m.
\end{equation}
(The argument $n-1$ is suppressed on the right hand side of the second
equality.)
The average in Eq.\ (\ref{eq:Tgeneral}) is the full disorder 
average, applied to all cavities in the array. Taking the limit
$g_{\rm c} \to \infty$ at fixed $n/g_{\rm c}$, Eq.\ (\ref{eq:Tgeneral})
is mapped to a coupled
set of differential equations for the moments of ${\cal
T}^{\rm q}$ which is identical to the corresponding set of
differential equations for a disordered
wire.\cite{kn:tartakovski1995,kn:brouwer1998b}
A subset of these equations can be resummed into a 
partial differential equation for the generating function $F_{\beta}$,
with\cite{kn:rejaei1996,kn:brouwer1996b}
\begin{eqnarray}
F_1
%  (\theta_1,\theta_2,\theta_3) 
 &=&
\left\langle \det \prod_{\pm}
 \left( \frac{2 + (\cos(\theta_3)-1) {\cal T}^{\rm q}}{2 +
(\cosh(\theta_1\pm\theta_2)-1) {\cal T}^{\rm q}}
 \right)^{1/2} \right\rangle, \nonumber \\
F_2
%  (\theta_1,\theta_3) 
 &=& 
 \left\langle 
\det \left( \frac{2 + (\cos(\theta_3)-1) {\cal T}^{\rm q}}{2 +
(\cosh(\theta_1)-1) {\cal T}^{\rm q}} \right) \right\rangle.
\end{eqnarray}
%which reads
%\begin{eqnarray}
%  \frac{\xi}{2} \frac{\partial}{\partial L} F &=&
%  \sum_{j=1,3}
%  \frac{1}{J(\theta_1,\theta_3)}
%  \frac{\partial}{\partial \theta_j}
%  J(\theta_1,\theta_3)
%  \frac{\partial}{\partial \theta_j}
%  F, \label{eq:Fevol}
%\end{eqnarray}
%with $L/\xi = n/2 g_{\rm c}$ and
%\begin{equation}
%  J(\theta_1,\theta_3) = \frac{\sin(\theta_3) \sinh(\theta_1)}{(\cosh(\theta_1)-\cos(\theta_3))^2}.
%\end{equation}
As shown in Refs.\ \onlinecite{kn:rejaei1996,kn:brouwer1996b}, 
the resulting theory of localization in quasi-one dimension
is formally equivalent to that obtained from the
one-dimensional nonlinear sigma model.\cite{kn:efetov1983,kn:efetov1983b,kn:zirnbauer1992,kn:mirlin1994}

\section{Semiclassical formalism}
\label{sec:2}

%\texttt{It would be nice, if we could stick to the notation convention $q\to$
%  coordinates, $p\to$ momenta. Would you mind?}

The central object in the trajectory-based semiclassical theory of
localization in an array of ballistic chaotic cavities is a 
semiclassical representation 
of the scattering matrix $S^{\rm q}$. In the semiclassical
representation, the discrete transverse momenta become
continuous variables, so that the scattering matrix $S^{\rm q}$ becomes a 
`scattering kernel' $S_{ij}(\pout,\pin)$. Following 
standard semiclassical approximations, this 
scattering kernel 
is then represented as a sum over classical trajectories $\alpha$ 
connecting contact $j$ to contact $i$,\cite{kn:jalabert1990,kn:stone1995}
\begin{equation}
S_{ij}(\pout,\pin) = \frac{1}{\sqrt{2 \pi \hbar}} \sum_{\alpha} A_{\alpha}
e^{i {\cal S}_{\alpha}/\hbar},
\label{eq:Ssemi}
\end{equation}
such that the transverse momentum of $\alpha$ upon entrance and exit
equals $\pin$ and $\pout$, respectively. Further, ${\cal S}_{\alpha}$ is the
classical action of $\alpha$, and $A_{\alpha}$ the stability amplitude,
\begin{equation}
A = \left| \frac{\partial \pout}{\partial q}
\right|^{-1/2},
\end{equation}
where $q$ is the transverse position upon entrance into the
sample. Maslov indices and other additional phase shifts are included 
in ${\cal
S}_{\alpha}$. Because the transverse modes in the quantum mechanical
formulation are linked to the absolute value of the transverse
momentum, not to the transverse momentum itself,
the semiclassical counterpart of the products $S_{ij}
S_{kj}^{\dagger}$ and $S_{jk}^{\dagger} S_{ji}$ 
consist of two contributions: one in which
transverse momenta in contact $j$ are equal and one in which the
transverse momenta are opposite,
\begin{eqnarray}
[S_{ij}^{\vphantom{M}} S_{kj}^{\dagger}](\pout,\pin) &=&
 \sum_{\pm} 
\int_{-p_F}^{p_F} \frac{dp''}{2 \pi \hbar}
%  \nonumber \\ && \mbox{} \times
  S_{ij}^{\vphantom{M}}(\pout,p'') S_{kj}^{\dagger}(\pin, \pm p''),
% + S_{ij}^{\vphantom{M}}(\pout,\pin) S_{kj}^{\dagger}(\pout,- p'')],
 \nonumber
\end{eqnarray}
\begin{eqnarray}
 [ 
 S_{jk}^{\dagger} 
 S_{ji}^{\vphantom{M}}]
 (\pin',\pin) 
 &=&
 \sum_{\pm} 
\int_{-p_F}^{p_F} \frac{dp''}{2 \pi \hbar}
%  \nonumber \\ && \mbox{} \times
 S_{jk}^{\dagger}(\pm p'',\pin') S_{ji}^{\vphantom{M}}(p'',\pin),
% +
%  S_{jk}^{\dagger}(-\pout,\pin') S_{ji}^{\vphantom{M}}(\pout,\pin)],
 \nonumber \\
\label{eq:Sproduct}
\end{eqnarray}
Together, Eqs.\ (\ref{eq:Ssemi}) and (\ref{eq:Sproduct}) specify how
products of the quantum-mechanical scattering matrix and its
hermitian conjugate are expressed in terms of classical trajectories.
[The ``$-$'' terms in the summations were omitted from the
semiclassical expression for the conductance, Eq.\ (\ref{eq:Gsemi})
above.]

For a theory of Anderson localization, we are interested in the
trace of a product of alternating factors $S$ and $S^{\dagger}$ 
or in the product of such traces. 
%ensemble averages of products of traces of the form
%\begin{equation}
%  X_{i_1\ldots i_{2m}} =
%  \tr S_{i_1 i_2} S_{i_3 i_2}^{\dagger} \ldots
%  S_{i_1 i_{2m}}^{\dagger}.
%\end{equation}
%Here each factor $S_{ij}$ is represented in terms of classical
%trajectories as in Eq.\ (\ref{eq:Ssemi}) above, and the ``product'' 
%$S_{i_1 i_2} S_{i_3 i_2}^{\dagger}$ is defined as
%\begin{eqnarray}
%  [S_{i_1 i_2} S_{i_3 i_2}^{\dagger}](p,q) &=&  \frac{1}{{2 \pi \hbar}} 
%  \sum_{\sigma' = \pm 1} \int_{-p_F}^{p_F} dp'
%  \nonumber \\ && \mbox{} \times
%  S_{i_1 i_2}(p,p') S_{i_3 i_2}^{\dagger}(p,\sigma' p').
%  \label{eq:Sproduct}
%\end{eqnarray}
Using the semiclassical representation (\ref{eq:Ssemi}),
%for $S$ and its complex conjugate for $S^{\dagger}$, 
a polynomial function $F(S,S^{\dagger})$ that involves the alternating
product of $n$ factors 
$S$ and $n$ factors
$S^{\dagger}$ is written as a summation over $2n$
classical trajectories $\alpha_1,\ldots,\alpha_n$ and
$\beta_1,\ldots,\beta_n$, one trajectory for each factor $S$ or
$S^{\dagger}$, respectively. Each configuration of classical trajectories
is weighed by a phase factor $\exp(i \Delta {\cal S}/\hbar)$ with
\begin{equation}
\Delta {\cal S} = \sum_{i=1}^{n} {\cal S}_{\alpha_i} - \sum_{i=1}^{n} {\cal
S}_{\beta_i}.
\label{eq:deltaS}
\end{equation}
Building on work by Richter and 
Sieber,\cite{kn:sieber2001,kn:sieber2002,kn:richter2002}
Haake and coworkers have identified a hierarchy of families of classical
trajectories $\alpha_1$,\ldots,$\beta_n$ that contribute to the
average $\langle F 
\rangle$,\cite{kn:mueller2004,kn:mueller2005,kn:heusler2006,kn:mueller2007}
where the average is taken with respect to variations of the Fermi
energy while keeping the classical dynamics (i.e., the shape of the
cavities) fixed.
Their identification is based on the recognition that families of
trajectories contribute to $\langle F \rangle$ only if their
total action difference $\Delta {\cal S}$ is of order $\hbar$ 
systematically, which happens only if the trajectories $\alpha_i$ 
are piecewise and pairwise identical to the
trajectories $\beta_j$, $i,j=1,\ldots,n$, up to classical phase space 
distances of order ${\hbar}^{1/2}$.\cite{kn:sieber2001,kn:sieber2002}
Trajectories that are separated by larger phase space distances have
typical action differences $\Delta {\cal S}$
that are parametrically larger than
$\hbar$, so that their contribution vanishes
upon taking the average. 

The simplest choice for a family of trajectories for which $\Delta
{\cal S}$ is of order $\hbar$ systematically is if each
trajectory $\alpha_i$ equals another trajectory $\beta_j$ for the full
length of the trajectory. Calculating $\langle F \rangle$ from 
the contribution from
such families of trajectories only
is known as the ``diagonal approximation''.\cite{kn:jalabert1990,kn:baranger1993,kn:baranger1993b}
Nontrivial families of trajectories
emerge from the possibility of small-angle encounters between 
trajectories, at which more than two trajectories are within a
phase space distance $\sim \hbar^{1/2}$.\cite{kn:aleiner1996}
At such small-angle
encounters, the pairing between the $\alpha_i$ and $\beta_j$ can be
changed --- so that now trajectories need to be piecewise
identical only. The duration of a small-angle encounter is the
``Ehrenfest time'' $\tau_{\rm E} = \lambda^{-1} \ln(p_F l/\hbar)$,
where $\lambda$ is the Lyapunov exponent of the classical dynamics,
$p_F$ the Fermi momentum, and $l$ a characteristic length scale of the
classical dynamics. The fundamental 
action integrals corresponding to each small-angle encounter are known,
\cite{kn:spehner2003,kn:turek2003,kn:mueller2004,kn:mueller2005} and
the resulting theory takes the form of  
simple combinatorial rules with which any product of traces of
products the scattering matrix and its hermitian conjugate can be 
calculated to arbitrary order in $\hbar$ from the semiclassical 
representation of $S$, provided that the 
Ehrenfest time $\tau_{\rm E}$ be much smaller than the sample's mean
dwell time $\tau_{\rm D}$.\cite{kn:mueller2004,kn:mueller2005,kn:heusler2006,kn:mueller2007}
(The case of finite $\tau_{\rm E}/\tau_{\rm D}$ is considerably more
complicated, see, {\em e.g.}, Refs.\
\onlinecite{kn:whitney2006,kn:rahav2006,kn:brouwer2006,kn:brouwer2007b},
but not relevant for a semiclassical theory of Anderson localization.)

In the remainder of this text, we refer to a calculation of the
energy-average $\langle F \rangle$ using contributions from families
of trajectories thus constructed as the ``trajectory-based
semiclassical formalism''.
Although there is no formal proof that this formalism is exact,
{\em i.e.}, that there are no other
contributions to $\langle F \rangle$ than from families of 
piecewise paired classical trajectories, the formalism satisfies all
known conservation rules and calculations based on 
trajectory-based semiclassics have been found to agree with fully
quantum mechanical calculations whenever
applicable.\cite{kn:mueller2007,kn:brouwer2007b} The present calculation can be viewed as
another demonstration of the validity of 
trajectory-based semiclassics, by showing that
the same formalism can serve as the starting point of a theory of
localization.

%Here we show that one is able to reproduce the existing theories of
%localization in quasi one dimension if one considers the contributions
%to $F$ from the same families of trajectories.

While we do not need the detailed results of the trajectory-based
semiclassical formalism, 
there are two properties of ensemble averages
calculated using that formalism that are particularly relevant for our 
calculations below:\\ 
%(i) All
%contributions from $\sigma=-1$ in the definition (\ref{eq:Sproduct})
%of the product of two scattering kernels vanish after taking the
%ensemble average. 
(i) All averages are compatible with
the condition of unitarity,\cite{kn:mueller2007}
\begin{eqnarray}
\sum_{j} [S_{ij}^{\vphantom{M}} S^{\dagger}_{kj}](p,p') &=&
\sum_{j} [S^{\dagger}_{jk} S_{ji}^{\vphantom{M}}](p,p') 
 \nonumber \\ &=& \delta_{ik} \delta(p-p').
\end{eqnarray}
(ii) For a product $S_{ij}^{\vphantom{M}} S^{\dagger}_{kj}$ or $S^{\dagger}_{ji}
S_{jk}^{\vphantom{M}}$, the trajectory $\alpha$ of the semiclassical
representation for $S$ and the trajectory $\beta$ of the semiclassical 
representation of $S^{\dagger}$ at contact $j$ satisfy 
\begin{equation}
p_{\alpha} = p_{\beta}, \ \ |q_{\alpha} - q_{\beta}| \lesssim
\hbar/p_F.
\end{equation}
In particular, this implies that there is no contribution from the
second term in Eq.\ (\ref{eq:Sproduct}) for a product of two
scattering kernels.\cite{kn:whitney2006,kn:rahav2006}\\
%Whereas the properties (i) and (ii) are satisfied for the average
%of {\em any} product of traces of products of alternating factors $S$
%and $S^{\dagger}$ to arbitrary order in $\hbar$, they have not been 
%shown to follow from the semiclassical expression (\ref{eq:Ssemi})
%for $S$ {\em before} averaging. However, since our goal is a
%statistical theory of the transport --- all our statements will refer
%to averaged quantities only --- we will accept these 
%two properties on the level of the sample-specific semiclassical 
%scattering kernel in the arguments that follow below.

\section{Array of ballistic cavities}
\label{sec:semi}

In the perturbative regime, the semiclassical theories of quantum
corrections to transport and to the density of states closely
followed the corresponding theory for disordered metals. The
semiclassical formalism described in the previous section was
instrumental in formalizing
the relation between the two types of theories. Motivated by this
correspondence, we now look for the possibility to adapt the theory of
localization in an array of disordered cavities [Sec.\ \ref{sec:dis}] to
the case of an array of ballistic cavities.

Thus, paraphrasing the arguments of Sec.\ \ref{sec:dis}, the goal of
our calculation will be to find the full probability distribution of
the function
\begin{equation}
{\cal T}(n;\pout,\pin) = [S_{12} S_{12}^{\dagger}](\pout,\pin)
\end{equation}
for an array of $n$ cavities. As shown in Sec.\ \ref{sec:dis}, there
are two ways in which this can be accomplished:
\begin{enumerate}
\item Using a stochastic approach, in which one considers the
 stochastic evolution of the function ${\cal T}(n;\pout,\pin)$ as a function
 of $n$, or through
\item the construction of a set of recursion relations for all moments
 of ${\cal T}(n)$.
\end{enumerate}
In both cases, the resulting theory is formally identical to the known 
theories of localization in quasi-one dimension.

Although technically simpler, the stochastic approach is at odds with
the goals of a semiclassical theory for ballistic localization: The
goal of a theory of localization in an array of ballistic cavities is
to describe an array of cavities with a fixed shape, using variations
of the Fermi energy as the only source of statistical
fluctuations. Since the Fermi energy is set globally, for all cavities
at the same time, quantum corrections for different cavities are not
independent, and a stochastic approach is ruled out a priori. A
stochastic approach is possible, however, if one relaxes the goals of
the theory, allowing for small variations of the shape of each cavity,
or for variations of a ``gate voltage'' that sets the Fermi energy of
each cavity independently.

Below we first describe the stochastic approach.
%, the construction of a
%stochastic recursion relation for ${\cal T}$ for the case that the
%Fermi energy of each cavity can be varied independently. 
In Sec.\
\ref{sec:b2} we
consider the case of broken time-reversal symmetry, which is
technically simpler. The discussion of localization in the
presence of time-reversal symmetry is given in Sec.\ \ref{sec:time}.
In Sec.\ \ref{sec:hierarchy}
we discuss how a hierarchy of recursion relations for the moments of 
${\cal T}$ can be constructed, where the average is taken with respect
to variations of the Fermi energy of the entire array only.

%An additional advantage of this alternative
%approach is that it considers energy averaged quantities
%only, for which the two properties (i) and (ii) above 
%have been shown to follow directly from the semiclassical averaging
%scheme \cite{kn:heusler2006,kn:mueller2007}.

\subsection{Stochastic approach}
\label{sec:b2}

The stochastic approach deals with (statistical) properties of the
function ${\cal T}(n)$ before averaging.
Although the properties (i) and (ii) of the trajectory-based
semiclassical formalism [Sec.\ \ref{sec:2}] are satisfied for the 
average of {\em any} product of traces of products alternating factors $S$
and $S^{\dagger}$ to
arbitrary order in $\hbar$, they have not been 
shown to follow from the semiclassical scattering matrix (\ref{eq:Ssemi})
{\em before} averaging. However, since our goal is a
statistical theory of the transport --- the final statements of the
theory will refer
to averaged quantities only --- we will accept these 
two properties on the level of the sample-specific semiclassical 
scattering kernel $S(n;\pout,\pin)$ in the arguments that follow below.

\begin{figure}
\includegraphics[width=0.5\hsize]{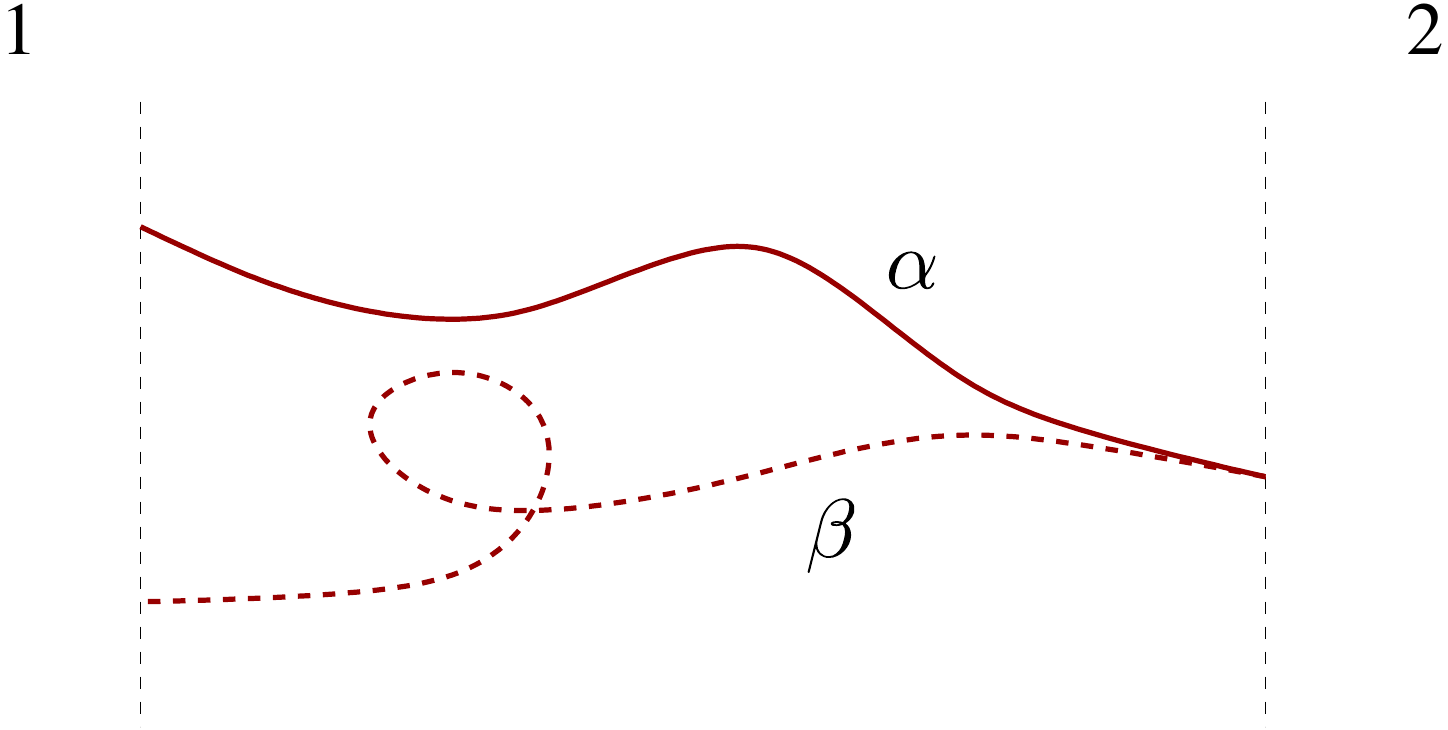}
\caption{\label{fig:2} (Color online)
Example of a pair of trajectories $\alpha$ and $\beta$
contributing to ${\cal T}(\pout,\pin) = [S_{12} S_{12}^{\dagger}](\pout,\pin)$. 
Upon entering the sample, the two
trajectories have different transverse momenta $p$ and $q$, and may
enter at different transverse positions. Upon exiting the sample,
$\alpha$ and $\beta$ have the same transverse momentum, and exit at
transverse positions a distance $\lesssim \hbar/p_F$ apart.}
\end{figure}

Starting from Eq.\ (\ref{eq:Ssemi}), 
the kernel ${\cal T}$ is expressed as a double sum
over classical trajectories $\alpha$ and $\beta$ that connect the
entrance and exit contacts [Fig.\ \ref{fig:2}],
\begin{equation}
{\cal T}(\pout,\pin) = \sum_{\alpha,\beta} \frac{A_{\alpha} A_{\beta}}{2 \pi
\hbar} e^{i({\cal S}_{\alpha} - {\cal S}_{\beta})/\hbar}.
\end{equation}
Here $p$ and $q$ are the transverse momenta of $\alpha$ and $\beta$
upon entrance. The two trajectories have equal transverse momenta upon 
exit, and exit at positions a quantum uncertainty $\sim \hbar/p_F$ 
apart --- see
property (ii) above.  Below, we 
express the difference $\delta {\cal T} = {\cal T}(n) - {\cal T}(n-1)$ 
in terms of classical trajectories. We first calculate the average
$\langle \delta {\cal T} \rangle_{n}$, where the average is taken with
respect to variations of the Fermi energy of the $n$th cavity (or of
its shape), while
keeping the Fermi energy and shape of the other cavities fixed. After
that, we calculate the variance of $\delta {\cal T}$ and the
higher cumulants. 
%Sections \ref{sec:3a}--\ref{sec:3c} deal with the
%technically simpler case of broken time-reversal
%symmetry. Modifications in the presence of time-reversal symmetry are
%discussed in Sec.\ \ref{sec:3d}.

%\subsection{Average of $\delta {\cal T}$}
%\label{sec:3a}

\begin{figure}
\includegraphics[width=\hsize]{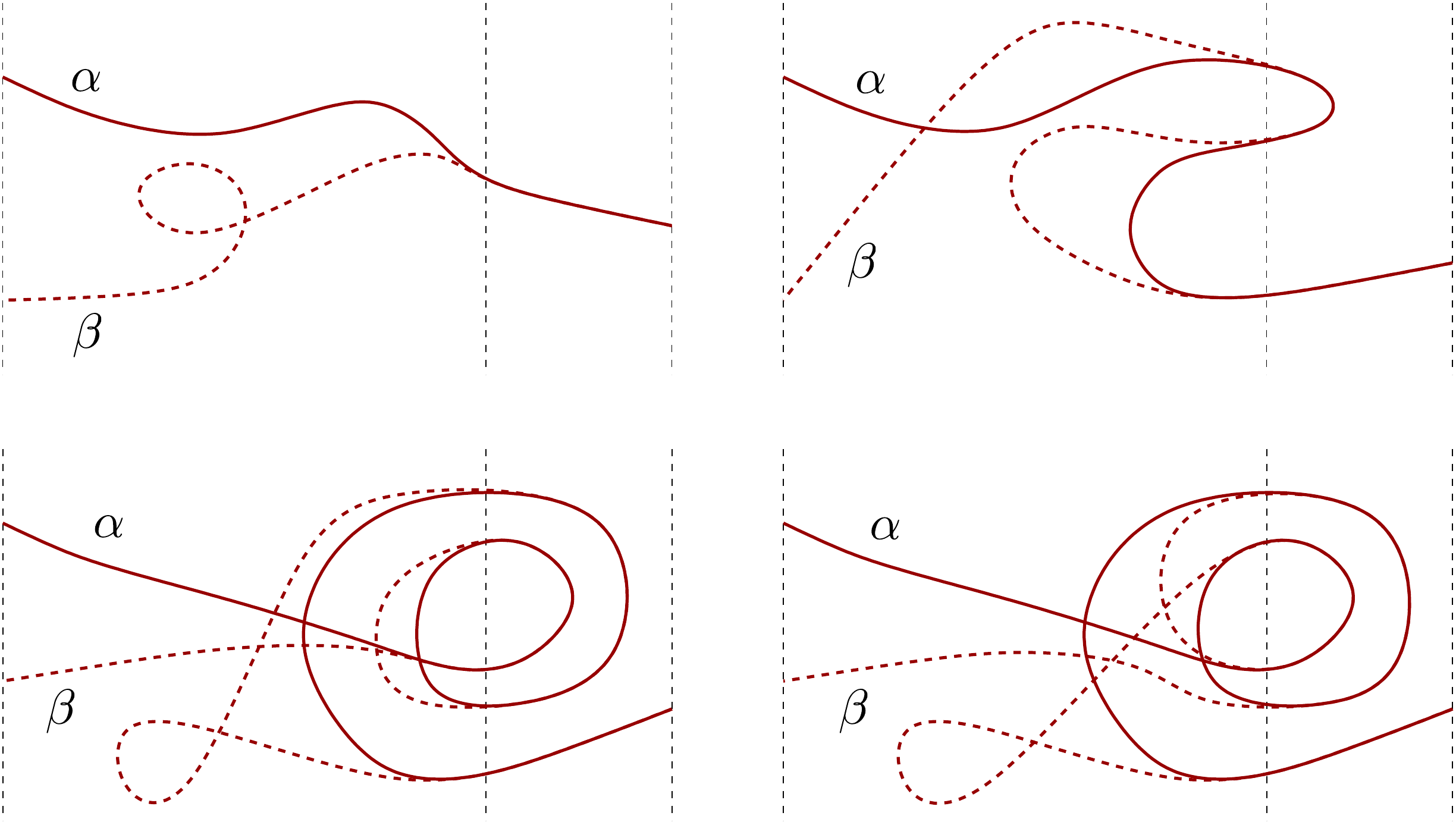}
\caption{\label{fig:3} (Color online)
Examples of trajectories $\alpha$ and $\beta$
contributing to $\langle {\cal
T}(n)\rangle_{n}$, where an average is taken over the Fermi energy
of the $n$th cavity only. The four panels show
trajectories contributing for $m=1$ (top left), $m=2$ (top right),
$m=3$ with diagonal pairing (bottom left), and $m=3$ with non-diagonal
pairing (bottom right). The number $m$ counts the number of times
$\alpha$ and $\beta$ enter the $n$th cavity from the $(n-1)$st
cavity. The three dashed lines in each panel represent the entrance
contact, the contact between the $(n-1)$st cavity and the $n$th
cavity, and the exit contact (from left to right). In the $n$th
cavity, the trajectories $\alpha$ and $\beta$ are piecewise equal up
to quantum uncertainties.}
\end{figure}

{\em Average of $\delta {\cal T}$.}
When calculating the average $\langle \delta {\cal T} \rangle_{n} =
\langle {\cal T}(n) \rangle_{n} - {\cal T}(n-1)$ to leading order in
$1/g_{\rm c}$ in the absence of time-reversal symmetry, it will be
sufficient to calculate $\langle {\cal T}(n) \rangle_{n}$ in 
the diagonal approximation by considering trajectories $\alpha$ and 
$\beta$ that are ``paired'' in the $n$th cavity. Note, however,
that trajectories need not be paired in the first $n-1$
cavities, because no average is taken there. Each trajectory
is classified by the
number of times $m_{\alpha}$, $m_{\beta}$ that it enters the
$n$th cavity from the $(n-1)$th cavity. Hence, for each trajectory 
there are $m_{\alpha}$ and $m_{\beta}$ segments in the $n$th cavity, 
which we label as $\alpha_1$, \ldots, $\alpha_{m_{\alpha}}$ and
$\beta_1$, \ldots, $\beta_{m_{\beta}}$. Since trajectories are paired
in the $n$th cavity, $m_{\alpha} = m_{\beta} = m$. Examples
of trajectory pairs $\alpha$ and $\beta$ with $m=1$, $2$, and $3$ are
shown in Fig.\ \ref{fig:3}. 
Since trajectories need to be paired upon exit, $\alpha_m$ has to be 
paired with $\beta_m$. While there are $(m-1)!$ ways in which the
remaining segments can be paired, we now show that the ``diagonal
pairing'',  
$\alpha_j$ paired with $\beta_j$, $j=1,\ldots,m$, gives the main
contribution to $\langle \delta {\cal T} \rangle_{n}$, whereas
all other pairings give contributions a factor $1/g_{\rm c}$ 
smaller. 

Cutting the trajectories $\alpha$ and $\beta$ at the contact between
the $(n-1)$st and $n$th cavity also separates the part of each
trajectory that resides in the first $n-1$ cavities into $m$
segments. The first segment of each trajectory connects the entrance
contact to the exit of the $(n-1)$st cavity; 
All other segments connect the exit contact of the $(n-1)$st cavity
to itself. Since the electron dynamics in the $n$th cavity is
fully ergodic, the
positions and transverse momenta with which these segments cross the
interface between the $(n-1)$st and $n$th cavity are fully
random, without correlations between different segments. [Correlations
can be ruled out down to quantum phase space distances $\sim 
\hbar/(p_F L)^{1/2}$ because
the dwell time $\tau_{\rm D} \gg \tau_{\rm E}$.] Hence,
these $m$ segments can be interpreted as the semiclassical
representation of a product of $S_{12}$, $S_{12}^{\dagger}$, and $m-1$ 
factors $S_{22}$ and $S_{22}^{\dagger}$, {\em before} taking any
average. Note that, although trajectories 
$\alpha,\beta$ that are ``paired'' in the $n$th cavity
have a phase space distance of order $\hbar/(p_F L)^{1/2}$
or smaller when they enter/exit the $n$th cavity from/to the $(n-1)$st
cavity, such trajectory pairs are sufficient for the semiclassical
calculation of the complete kernels ${\cal T}$ because 
of property (ii) of Sec.\ \ref{sec:2}. 

For the diagonal pairing, the $m$ segments of
$\alpha$ and $\beta$ that reside in the first $n-1$ cavities generate a 
particularly simple product of factors $S$ and $S^{\dagger}$, 
${\cal T}(n-1) [\mbox{tr}\, {\cal R}(n-1)]^{m-1}$, where
\begin{equation}
{\cal R}(n;p,p') = [S_{22}(n)^{\dagger} S_{22}(n)](p,p')
\end{equation}
is the reflection coefficient for the first $n-1$
cavities, seen from the exit contact of the $(n-1)$st cavity, and
\begin{equation}
\mbox{tr}\, {\cal R}(n) =
\int_{-p_F}^{p_F} dp\, {\cal R}(n;p,p).
\end{equation}
Using the trajectory-based semiclassical formalism to evaluate the diagonal
trajectory sums in the $n$th 
cavity,\cite{kn:baranger1993,kn:baranger1993b,kn:richter2002,kn:heusler2006,kn:mueller2007} one then
easily finds that the diagonal pairing gives
\begin{eqnarray}
\langle {\cal T}(n) \rangle_{n}
&=& \sum_{m=1}^{\infty} \frac{1}{2^{m} g_{\rm c}^{m-1}} {\cal T}(n-1)
[\mbox{tr}\, {\cal R}(n-1)]^{m-1}
\nonumber \\ &=&
\frac{{\cal T}(n-1) g_{\rm c}}{2 g_{\rm c} - \mbox{tr}\, {\cal R}(n-1)}.
\label{eq:X1p}
\end{eqnarray}
Using unitarity, one has
\begin{equation}
\mbox{tr}\, {\cal R}(n) = g_{\rm c} - \mbox{tr}\, {\cal T}(n),
\end{equation}
so that Eq.\ (\ref{eq:X1p}) can be rewritten as
\begin{eqnarray}
\langle \delta {\cal T} \rangle_{n}
&=& - \frac{1}{g_{\rm c}} {\cal T}(n-1) \mbox{tr}\, {\cal T}(n-1) 
,
\label{eq:ttdiffavg}
\end{eqnarray}
up to corrections of order $1/g_{\rm c}^2$.
This is precisely the semiclassical equivalent of the recursion
relation of Eq.\ (\ref{eq:ttdiff}), averaged over the Fermi energy or
shape of
the $n$th cavity.

It remains to show that non-diagonal pairings, in which a segment
$\alpha_j$ is not paired with $\beta_j$ give a contribution to
$\langle \delta {\cal T} \rangle_{n}$ that can be neglected in the
limit of large $g_{\rm c}$. Hereto we first consider the case $m=3$, 
for which the only possible non-diagonal
pairing is $\alpha_1 \leftrightarrow \beta_2$, $\alpha_2
\leftrightarrow \beta_1$ [Fig.\ \ref{fig:3}, bottom right]. 
For this pairing, the three segments of the
trajectories that reside in the first $n-1$ cavities generate the
semiclassical representation of a product of six scattering matrices,
$S_{12}^{\vphantom{M}} S^{\dagger}_{22} S_{22}^{\vphantom{M}} S^{\dagger}_{22} S_{22}^{\vphantom{M}}
S_{12}^{\dagger}$. From unitarity, one has
\begin{eqnarray}
S_{12}^{\vphantom{M}} S^{\dagger}_{22} S_{22}^{\vphantom{M}} S^{\dagger}_{22}
S_{22}^{\vphantom{M}} S_{12}^{\dagger}
&=&
{\cal T}(n-1) - 2 {\cal T}(n-1)^2 
 \nonumber \\ && \mbox{}
 + {\cal T}(n-1)^3.
\end{eqnarray}
For comparison, the diagonal pairing for $m=3$ generates ${\cal
T}(n-1)\,[\mbox{tr}\, {\cal R}(n-1)]^2$, which is a factor $\sim g_{\rm
c}^2$ larger because $\mbox{tr}\, {\cal R}(n-1) \sim g_{\rm c}$.
Using the ergodic dynamics in the $n$th cavity, the non-diagonal 
pairing of segments in the $n$th cavity for
$m=3$ gives a contribution to $\langle \delta {\cal T}
\rangle_{n}$ equal to
\begin{eqnarray}
\langle \delta {\cal T} \rangle_{n}^{(3,{\rm non\, diag})}
&=&
\frac{1}{8 g_{\rm c}^2} [{\cal T}(n-1) - 2 {\cal T}(n-1)^2 
 \nonumber \\ && \mbox{} + {\cal
T}(n-1)^3].
\end{eqnarray}
This is a factor $\sim 1/g_{\rm c}$ smaller than the leading
contribution (\ref{eq:ttdiffavg})
to $\langle \delta {\cal T} \rangle_{n}$.
%
%Hence, unlike in the calculation of the diagonal pairing contribution
%to $\langle \delta {\cal T} \rangle$, for which each additional return 
%to the $(n-1)$st cavity gave an additional factor $\mbox{tr}\, 
%{\cal R}(n-1) \sim g_{\rm c}$, in this case no factors $g_{\rm c}$ are
%generated. Instead, one finds that 
%Hence, using the ergodic dynamics in the $n$th cavity, one finds 
%that the non-diagonal pairing of segments in the $n$th cavity for
%$m=3$ gives a contribution to $\langle \delta {\cal T}
%\rangle_{n}$ equal to
%\begin{eqnarray}
%  \langle \delta {\cal T} \rangle_{n}^{(3,{\rm non\, diag})}
%  &=&
%  \frac{1}{4 g_{\rm c}^2} {\cal F}'(n-1).
%\end{eqnarray}
%Comparing to Eq.\ (\ref{eq:ttdiffavg}), we see that this contribution 
%to $\langle \delta {\cal T} \rangle_{n}$ is
%a factor $1/g_{\rm c}$ smaller than the contribution to $\langle
%\delta {\cal T} \rangle_{n}$ that we found using the diagonal
%pairing. 

The same arguments can be used 
for $m > 3$: Non-diagonal pairings come at
the expense of at least two
factors $\mbox{tr}\, {\cal R}(n-1)$ and, hence, lead to
contributions to $\langle \delta {\cal T} \rangle_{n}$ that are at
least a factor $\sim 1/g_{\rm c}$ smaller than the contribution from 
diagonal pairing. These arguments can also
be used to show that contributions to 
$\langle {\cal T}(n) \rangle_{n}$ that involve
small-angle intersections of the trajectories in the $n$th cavity 
are a factor $1/g_{\rm c}$ smaller than the leading contribution
considered above.

%\subsection{Fluctuations of $\delta {\cal T}$}
%\label{sec:3b}

{\em Fluctuations of $\delta {\cal T}$.}
The fluctuations of $\delta {\cal T}$ are described by the covariance 
$\langle \delta {\cal T}(p_1,p_2) \delta {\cal T}(p_1',p_2') \rangle_{n}$.
We calculate $\langle \delta {\cal T}(p_1,p_2) \delta {\cal T}(p_1',p_2') 
\rangle_{n}$ from the identity 
\begin{eqnarray}
\langle \delta {\cal T}
\delta {\cal T}' \rangle_{n} &=&
%  \langle {\cal T}(n)^2 \rangle_{n} - 2 
%  \langle {\cal T}(n) {\cal T}(n-1) \rangle_{n} +
%  \langle {\cal T}(n-1)^2 \rangle_{n} \nonumber \\ &=&
\langle \delta ({\cal T} {\cal T}') \rangle_{n} 
- {\cal T}(n-1) \langle \delta {\cal T}' \rangle_{n}
 \nonumber \\ && \mbox{}
- {\cal T}'(n-1) \langle \delta {\cal T} \rangle_{n},
\label{eq:dtt}
\end{eqnarray}
where we used the shorthand notation ${\cal T}(n) = {\cal T}(n;p_1,p_2)$,
${\cal T}'(n)= {\cal T}(n;p_1',p_2')$, and $\delta ({\cal T} {\cal T}') =
{\cal T}(n) {\cal T}'(n) - {\cal T}(n-1) {\cal T}'(n-1)$.
The average $\langle \delta {\cal T} \rangle_{n}$ is
given by Eq.\ (\ref{eq:ttdiffavg}) above, so it remains to calculate
$\langle \delta ({\cal T} {\cal T}') \rangle$. The product ${\cal
T}(n) {\cal T'}(n)$ is represented as a sum over four classical
trajectories, $\alpha$, $\beta$, $\alpha'$, and $\beta'$. 
As before, we introduce the numbers 
$m_{\alpha}$, $m_{\beta}$, $m_{\alpha'}$ and $m_{\beta'}$ that
indicate how
often each trajectory enters the $n$th cavity. Since trajectories
always enter or exit the $n$th cavity in pairs, one has 
$m_{\alpha} + m_{\alpha'} = m_{\beta} + m_{\beta'}$. 

Unlike the average $\langle \delta {\cal T} \rangle_{n}$, for which
the only contribution came from the diagonal approximation in the
$n$th cavity with diagonal pairing of the segments $\alpha_j$ and
$\beta_j$, the average of the second moment $\langle (\delta {\cal T}
{\cal T}') \rangle_{n}$ has contributions from both non-diagonal
pairings in the diagonal approximation and from trajectories beyond
the diagonal approximation, which have small-angle encounters in the
$n$th cavity.  We first consider the diagonal approximation, for which
each segment $\alpha_i$ or $\alpha'_i$ is paired with another segment
$\beta_j$ or $\beta'_j$. For the diagonal pairing of segments,
$\alpha_j \leftrightarrow \beta_j$, $j=1,\ldots,m_{\alpha}=m_{\beta}$
and $\alpha'_j \leftrightarrow \beta'_j$, $j=1,\ldots,m_{\alpha'} =
m_{\beta'}$, we find
\begin{equation}
\langle \delta ({\cal T} {\cal T}') \rangle_{n}^{\rm diag}
= {\cal T}(n-1) \langle \delta {\cal T}' \rangle_{n}
+ {\cal T}'(n-1) \langle \delta {\cal T} \rangle_{n}.
\end{equation}
This contribution to $\langle \delta {\cal T} \delta {\cal T}'
\rangle_{n}$ precisely cancels the second and third terms in Eq.\
(\ref{eq:dtt}). Hence $\langle \delta {\cal T} 
\delta {\cal T}'
\rangle_{n}$ must be from non-diagonal pairings of the trajectory
segments within the diagonal approximation, or from trajectory
configurations beyond the diagonal approximation in the $n$th 
cavity. Since the latter class of trajectories have small-angle
encounters, we write
\begin{equation}
\langle \delta {\cal T} \delta {\cal T}'
\rangle_{n} =
\langle \delta ({\cal T} {\cal T}') \rangle_{n}^{\rm
non\ diag} +
\langle \delta ({\cal T} {\cal T}') \rangle_{n}^{\rm
enc}.
\end{equation}

\begin{figure}
\includegraphics[width=\hsize]{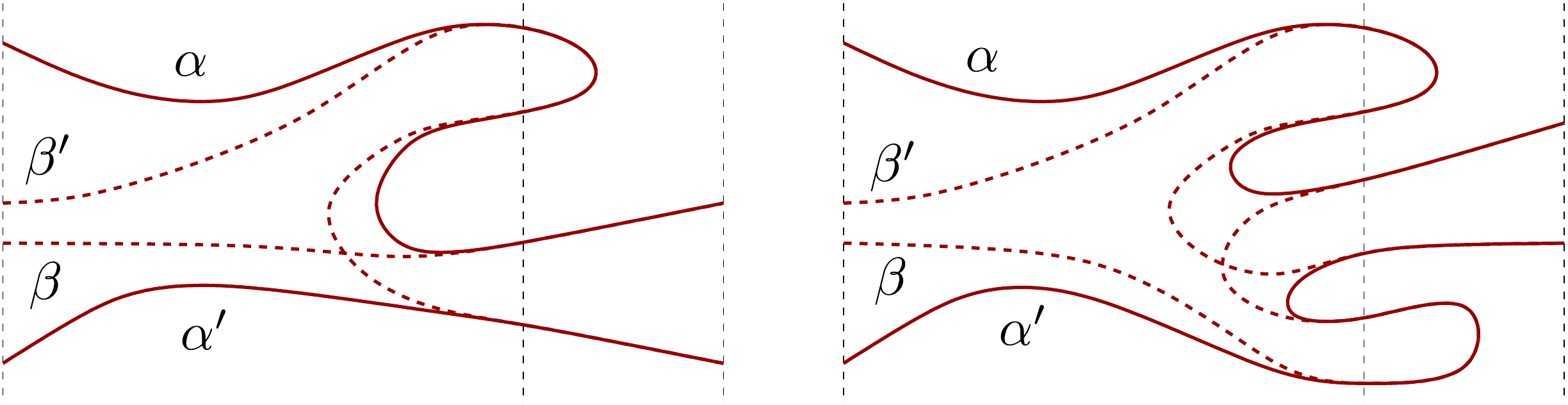}
\caption{\label{fig:4} (Color online)
Examples of four trajectories $\alpha$,
 $\beta$, $\alpha'$, and $\beta'$ contributing to $\langle {\cal
   T}(n;\pout,\pin) {\cal T}(n;\pout',\pin') \rangle_{n}$ for an array of chaotic
 cavities, where the average is taken with respect to the Fermi
 energy of the $n$th cavity only.  The examples shown here have a
 non-diagonal pairing with $m_{\alpha} = m_{\beta'} = 2$,
 $m_{\alpha'} = m_{\beta} = 1$ (left panel) and $m_{\alpha} =
 m_{\beta'} = m_{\alpha'} = m_{\beta} = 2$ (right panel).  In the
 $n$th cavity, the trajectories $\alpha$ and $\alpha'$ are piecewise
 equal to $\beta$ and $\beta'$, up to quantum uncertainties.}
\end{figure}

The leading non-diagonal pairing appears if one pairs the first $u$
segments of $\alpha$ with the first $u$ segments of $\beta'$, and the
last $m_{\alpha}-u$ segments of $\alpha$ with the last $m_{\alpha}-u$
segments of $\beta$, as well as the first $v$
segments of $\alpha'$ with the first $v$ segments of $\beta$, and the
last $m_{\alpha'}-v$ segments of $\alpha'$ with the last $m_{\alpha'}-v$
segments of $\beta'$, where $u=1,\ldots,m_{\alpha}-1$ and 
$v=1,\ldots,m_{\alpha}'-1$ are integers. 
%(Recall that $m_{\alpha} + m_{\alpha'} =
%m_{\beta} + m_{\beta'}$.) 
Two examples, with $m_{\alpha} = 2$,
$m_{\alpha'} = 1$, $u=1$, and $v=0$, and with $m_{\alpha} =
m_{\alpha'} = 2$ and $u=v=1$, are shown in Fig.\ \ref{fig:4}. 
%The
%contributions with $u=0$ or $v=0$ remain, all other contributions
%cancel against terms that arise from going beyond the diagonal
%approximation in the $n$th cavity. 
One then finds
\begin{eqnarray}
\langle \delta {\cal T} \delta {\cal T} \rangle^{\rm non\, diag}_{n}
&=& 
%  {\cal T}(p,q') {\cal F}(p',q)
%  \sum_{u=1}^{\infty} \sum_{m=u+1}^{\infty}
%  \sum_{m'=1}^{\infty}
%  \frac{1}{2^{m+m'} g_{\rm c}^{m+m'-2}}
%  (\mbox{tr}\, {\cal R})^{m+m'-3} 
%  \nonumber \\ && \mbox{} +
%  {\cal T}(p',q) {\cal F}(p,q')
%  \sum_{v=1}^{\infty} \sum_{m=1}^{\infty}
%  \sum_{m'=v+1}^{\infty}  
%  \frac{1}{2^{m+m'} g_{\rm c}^{m+m'-2}}
%  (\mbox{tr}\, {\cal R})^{m+m'-3} 
%  \nonumber \\ && \mbox{} +
%  {\cal T}(p',q) {\cal T}(p,q')
%  \sum_{u,v=1}^{\infty}
%  \sum_{m=u+1}^{\infty}
%  \sum_{m'=v+1}^{\infty}
%  \frac{(\mbox{tr}\, {\cal R}^2)}{2^{m+m'} g_{\rm c}^{m+m'-2}}
%  (\mbox{tr}\, {\cal R})^{m+m'-4}   
%  \nonumber \\ &=&
\frac{1}{g_{\rm c} }[{\cal F}(p_1',p_2) {\cal T}(p_1,p_2') 
\nonumber \\ && \mbox{}
+ {\cal F}(p_1,p_2') {\cal T}(p_1',p_2)
   \\ && \mbox{}
+ {\cal T}(p_1,p_2') {\cal T}(p_1',p_2)] 
% \nonumber \\ && \mbox{}
+
{\cal O}(1/g_{\rm c}^2), \nonumber
\end{eqnarray}
where
% $m = m_{\alpha}$, $m' = m_{\alpha'}$, 
${\cal T}(p_1,p_2) = {\cal
T}(n-1;p_1,p_2)$ and ${\cal F}(p_1,p_2) = {\cal F}(n-1;p_1,p_2)$, with
\begin{eqnarray}
{\cal F}(n) &=& \mbox{tr}\, S_{12} S^{\dagger}_{22} S_{22}
S^{\dagger}_{12}
\nonumber \\ &=&
{\cal T}(n) - {\cal T}(n)^2.
\end{eqnarray}
Other pairings give contributions of order $1/g_{\rm c}^2$ or
smaller.

\begin{figure}
\includegraphics[width=\hsize]{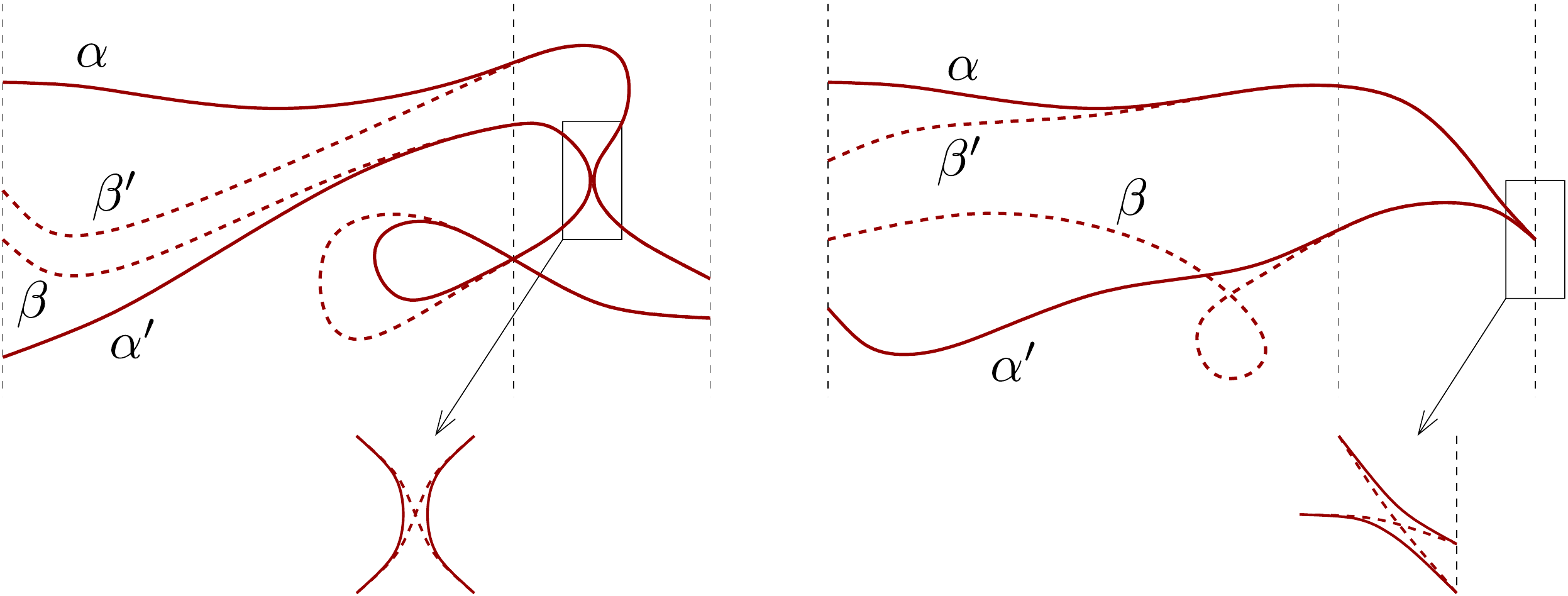}
\caption{\label{fig:5} (Color online)
Two examples of a set of four trajectories
contributing to $\langle {\cal T} {\cal T}' \rangle_{n}$ that have
a small-angle encounter in the $n$th cavity. For the left panel the
encounter is in the interior of the $n$th cavity; For the right
panel it touches the exit contact.}
\end{figure}

The second contribution to the fluctuations of $\delta {\cal T}$ comes
from trajectories with a small-angle encounter in the $n$th
cavity. Since we only need fluctuations of $\delta {\cal T}$ to
leading order in $1/g_{\rm c}$, it is sufficient to consider
trajectories with one small-angle encounter only.
Taking the small-angle encounter to be between the segments
$\alpha_{u}$,
$\beta'_{u}$, $\alpha'_{v}$, and $\beta'_{v}$, with $1 \le u,v \le
m,m'$, and pairing the first $u-1$ segments of $\alpha$ with the first 
$u-1$ segments of $\beta'$, and the
last $m-u$ segments of $\alpha$ with the last $m-u$ segments of
$\beta$, as well as the first $v-1$ segments of $\alpha'$ with the
first $v-1$ segments of $\beta$, and the last $m'-v$ segments of 
$\alpha'$ with the last $m'-v$ segments of $\beta'$, we find
\begin{widetext}
\begin{eqnarray}
\langle \delta {\cal T} \delta {\cal T} \rangle^{\rm enc}_{n}
&=& 
- {\cal T}(p_1,p_2') {\cal T}(p_1',p_2)
\sum_{u,v=1}^{\infty}
\sum_{m=u}^{\infty}
\sum_{m'=v}^{\infty}
\frac{(\mbox{tr}\, {\cal
R})^{m+m'-2}}{2^{m+m'+1}g_{\rm c}^{m+m'-1}} 
\nonumber \\ && \mbox{}
+ {\cal T}(p_1,p_2') {\cal T}(p_1',p_2)
\sum_{m=1}^{\infty}
\sum_{m'=1}^{\infty}
\frac{(\mbox{tr}\, {\cal
R})^{m+m'-2}  }{2^{m+m'}g_{\rm c}^{m+m'-1}} 
\nonumber \\ &=&
- \frac{1}{g_{\rm c}} {\cal T}(p_1,p_2') {\cal T}(p_1',p_2). 
\end{eqnarray}
\end{widetext}
where the first line comes from encounters in the interior of the
$n$th cavity and the second line comes from encounters that touch the
exit contact.\cite{kn:whitney2006,kn:brouwer2006c}
Examples of the two terms for $m=m'=1$ are shown in Fig.\
\ref{fig:5}.
[We do not need to consider encounters that touch the
contact between the $(n-1)$st cavity and the $n$th cavity, because the
contribution from such encounters is included in the (products) of the
kernel ${\cal T}$ of the first $n-1$ cavities.]

Combining everything we have
\begin{eqnarray}
\langle \delta {\cal T}(p_1,p_2) \delta {\cal T}(p_1',p_2') 
\rangle_{n}
&=& 
\frac{1}{g_{\rm c} }[{\cal F}(p_1',p_2) {\cal T}(p_1,p_2') 
 \nonumber \\ && \mbox{}
+ {\cal F}(p_1,p_2') {\cal T}(p_1',p_2)] 
\nonumber \\ && \mbox{}
+
{\cal O}(1/g_{\rm c}^2).
\label{eq:ddtresult}
\end{eqnarray}
Equation (\ref{eq:ddtresult}) is the semiclassical equivalent
of Eq.\ (\ref{eq:calTstat}).

Higher cumulants of $\delta {\cal T}$ can be calculated in the same
way. For the $k$th cumulant, one finds that only pairings that involve
trajectories out of all $k$ factors ${\cal T}$ contribute. Each
additional factor ${\cal T}$ involved in the pairing scheme
contributes an additional factor $1/g_{\rm c}$, which is why all
cumulants with $k > 2$ are of sub-leading order in $1/g_{\rm c}$.

%\subsection{Scaling theory of localization}
%\label{sec:3c}

Together, Eqs.\ (\ref{eq:ttdiffavg}) and (\ref{eq:ddtresult}) form the
semiclassical equivalent of the stochastic recursion relation
(\ref{eq:ttdiff}) used in the fully quantum mechanical theory of
localization in disordered quasi-one dimensional wires. In the limit
$g_{\rm c} \to \infty$ at fixed $n/g_{\rm c}$, Eqs.\
(\ref{eq:ttdiffavg}) and (\ref{eq:ddtresult}) can be mapped to
stochastic differential equation for the eigenvalues of ${\cal T}$,
which will be formally equivalent to the DMPK equation. 
%In this
%mapping, one makes the identification $L/\xi = n/2 g_{\rm c}$, where
%$L$ and $\xi$ are the length of the wire and the localization length,
%respectively \cite{kn:brouwer1996b}. 
The DMPK
equation, in turn, provides a full description of localization in
quasi-one dimensional wires.\cite{kn:dorokhov1982,kn:mello1988,kn:beenakker1997}

\subsection{Presence of time-reversal symmetry}
\label{sec:time}

In the presence of time-reversal symmetry both the average $\langle
\delta {\cal T} \rangle_{n}$ and the covariance $\langle \delta {\cal
T} \delta {\cal T}' \rangle_{n}$ are different. In both cases the
difference appears because one can pair time-reversed trajectories
when taking the ensemble average in the $n$th cavity. 

\begin{figure}
\includegraphics[width=\hsize]{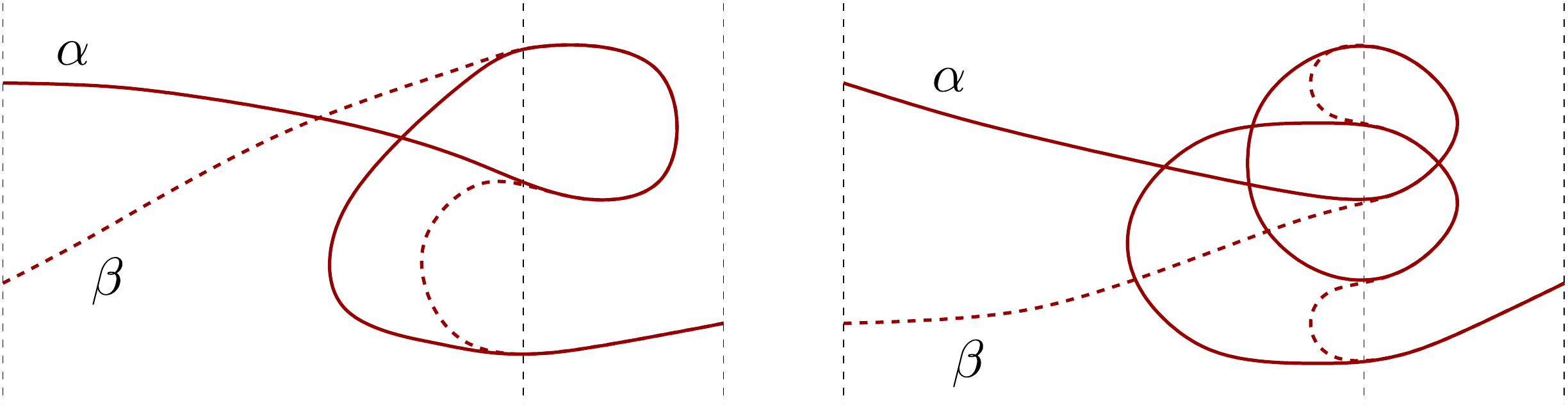}
\caption{\label{fig:6} (Color online)
Two examples of trajectory pairs contributing to
$\langle \delta {\cal T} \rangle_{n}$ in the presence of
time-reversal symmetry.}
\end{figure}

\begin{figure}
\includegraphics[width=\hsize]{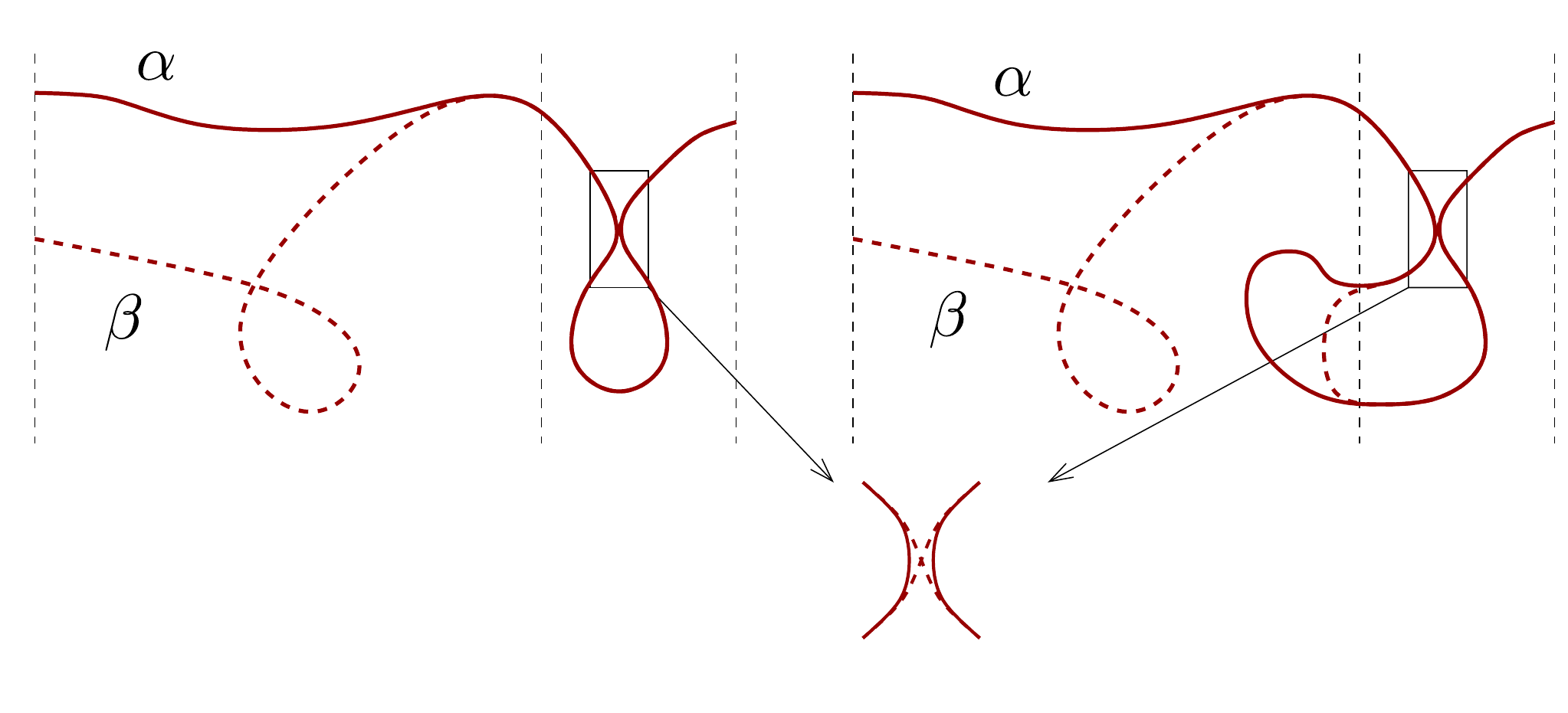}
%\epsffile{trajwlenc.eps}
\caption{\label{fig:7} (Color online)
Two examples of
 trajectory pairs contributing to $\langle \delta {\cal T}
 \rangle_{n}$ in the presence of time-reversal symmetry. The
 trajectory pairs have a small-angle encounter in the $n$th cavity.}
\end{figure}

There are two additional contributions to
the average $\langle \delta {\cal T} \rangle_{n}$. The first of these
arises from the diagonal approximation for the average in the $n$th
cavity. As before, we define the number $m = m_{\alpha} = m_{\beta}$
as the number of times the trajectories $\alpha$ and $\beta$ enter the
$n$th cavity from the $(n-1)$st cavity. The first additional
contribution to $\langle \delta {\cal T} \rangle_{n}$ then involves the 
pairing of segments $\alpha_{u+j}$ with the time-reversed of
$\beta_{u+v-j-1}$, $j=0,\ldots,v-1$, and $1 \le u \le u+v < m$,
\begin{eqnarray}
\langle \delta {\cal T} \rangle_{n}^{(1)} &=&
{\cal F}
\sum_{v=1}^{\infty}
\sum_{m=v+1}^{\infty}
\frac{1}{2^{m} g_{\rm c}^{m-1}} (\mbox{tr}\, {\cal R})^{m-2}
\nonumber \\ && \mbox{}
+
{\cal T}
\sum_{u=1}^{\infty}
\sum_{v=1}^{\infty}
\sum_{m=u+v+1}^{\infty}
\frac{\mbox{tr}\,{\cal R}^2}{2^{m} g_{\rm c}^{m-1}} (\mbox{tr}\, {\cal
R})^{m-3}
\nonumber \\ &=&
\frac{1}{g_{\rm c}}
\left( {\cal F} + {\cal T} \right).
\label{eq:tt1}
\end{eqnarray}
Examples for trajectory pairs contributing to the first and second
line in Eq.\ (\ref{eq:tt1}) are shown in Fig.\ \ref{fig:6}.
The second contribution comes from small-angle encounters inside $n$th
cavity involving the segments $\alpha_u$, $\beta_u$, $\alpha_{u+v}$,
and $\beta_{u+v}$, where $1 \le u \le u+v \le m$. For this contribution
one finds
\begin{eqnarray}
\langle \delta {\cal T} \rangle_{n}^{(2)} &=&
%  - {\cal T}
%  \sum_{u=1}^{\infty}
%  \sum_{m=u}^{\infty}
%  \frac{1}{2^{m+1} g_{\rm c}^{m}} (\mbox{tr}\, {\cal
%  R})^{m-1}  
%  \nonumber \\ && \mbox{}
%  - {\cal T}
%  \sum_{u=1}^{\infty}
%  \sum_{v=1}^{\infty}
%  \sum_{m=u+v}^{\infty}
%  \frac{1}{2^{m+1} g_{\rm c}^{m}} (\mbox{tr}\, {\cal
%  R})^{m-1}
%  \nonumber \\ &=&
- \frac{2}{g_{\rm c}} {\cal T}.
\end{eqnarray}
Examples of trajectory pairs contributing to $\langle \delta {\cal
T}^{(2)}\rangle_{n}$ are shown in Fig.\ \ref{fig:7}.
Using ${\cal F} = {\cal T} - {\cal T}^2$ and adding these two 
contributions to Eq.\ (\ref{eq:ttdiffavg}) one finds
\begin{eqnarray}
\langle \delta {\cal T} \rangle_{n}
&=& - \frac{1}{g_{\rm c}} {\cal T}(n-1) \mbox{tr}\, {\cal T}(n-1) 
- \frac{1}{g_{\rm c}} {\cal T}^2 ,
\label{eq:ttdiffavgb1}
\end{eqnarray}
up to corrections of order $1/g_{\rm c}^2$.  

For the fluctuations of $\delta {\cal T}$ in the presence of
time-reversal symmetry one finds an extra
contribution from pairing $\alpha_j$ with the time-reversed of
$\beta'_{u+1-j}$, $j=1,\ldots,u$, or $\alpha'_j$ with the
time-reversed of $\beta_{u+1-j}$, while following diagonal 
pairing rules for all other segments. Proceeding as before, we
find
\begin{eqnarray}
\langle \delta {\cal T}(p_1,p_2) \delta {\cal T}(p_1',p_2') 
\rangle_{n}
&=& 
\frac{1}{g_{\rm c} }[{\cal F}(p_1',p_2) {\cal T}(p_1,p_2') 
 \nonumber \\ && \mbox{}
+ {\cal F}(p_1,p_2') {\cal T}(p_1',p_2)
 \nonumber \\ && \mbox{}
+ 2 {\cal G}(p,p_1') {\cal G}^{\dagger}(p_2,p_2')] 
\nonumber \\ && \mbox{}
+
{\cal O}(1/g_{\rm c}^2).
\label{eq:ddtresultb1}
\end{eqnarray}  
where
\begin{eqnarray}
{\cal G}(p,p') &=& 
{\cal G}(p',p) \nonumber \\ &=& [S_{12} S_{22}^{\dagger}
S_{21}](p,p').
\end{eqnarray}
The stochastic process defined by Eqs.\ (\ref{eq:ttdiffavgb1})
and (\ref{eq:ddtresultb1}) precisely mirrors the stochastic process
(\ref{eq:ttdiff}) for the quantum mechanical matrix 
${\cal T}^{\rm q} = S_{12}^{\rm q}
S_{12}^{{\rm q}\dagger}$. 

\subsection{Recursion relations for the moments of ${\cal T}$}
\label{sec:hierarchy}

The direct construction of recursion relations for the moments of
${\cal T}$ is an alternative to the stochastic approach that avoids
extending the use of the properties (i) and (ii) of Sec.\ \ref{sec:2}
to sample-specific quantities and the necessity to define a
statistical ensemble by varying the Fermi energy or shape
of each cavity
individually. The construction of recursion relations for moments of
${\cal T}$ proceeds in the very same manner as the construction of the
stochastic recursion relations for ${\cal T}$, with the additional
requirement that trajectories are piecewise paired in all $n$
cavities, not only in the $n$th cavity. Since the
arguments of the preceding section did not rely on the structure of
the trajectories in the first $n-1$ cavities, one immediately
concludes that the recursion relations for the moments of ${\cal T}$
derived this way are identical to the recursion relations for the
moments of ${\cal T}$ one obtains from the stochastic
approach. 
Starting from the stochastic recursion relations
(\ref{eq:ttdiffavg}) and (\ref{eq:ddtresult}) or
(\ref{eq:ttdiffavgb1}) and (\ref{eq:ddtresultb1}) one arrives at 
the same hierarchy of recursion relations
(\ref{eq:Tgeneral}) derived for an array of disordered cavities. 

\begin{figure}
\includegraphics[width=\hsize]{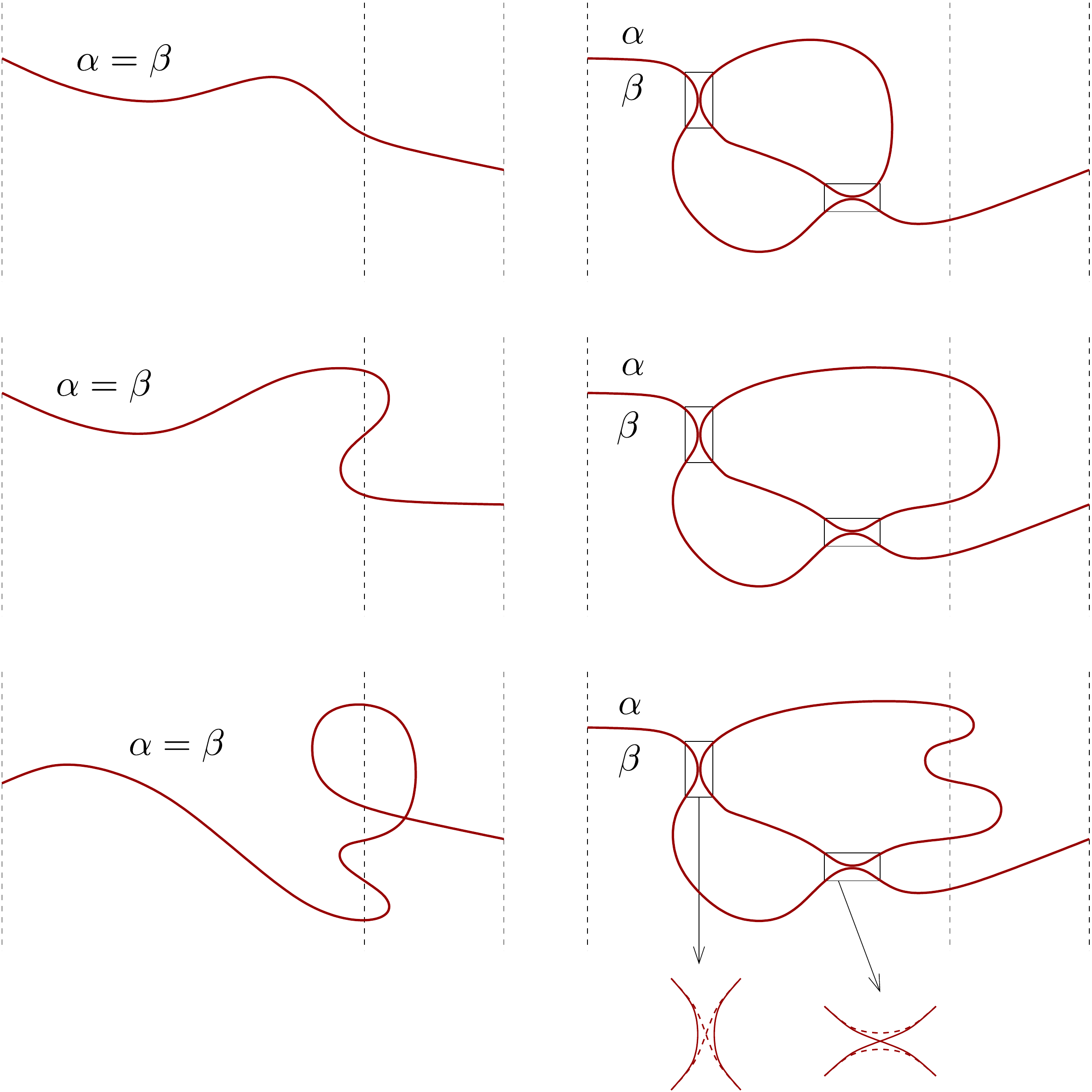}
\caption{\label{fig:avg} (Color online)
Examples of trajectory pairs contributing to
the full average $\langle {\cal T}(n) \rangle$ in the absence of
time-reversal symmetry. The trajectory pairs have $m=1$ (top), $m=2$
(center), and $m=3$ (bottom). The left panel shows
trajectory pairs without small-angle self intersections; The right
panel shows trajectory pairs with two small-angle self intersections
in the first $n-1$ cavities.}
\end{figure}

We illustrate this procedure for the recursion relation for
the first moment $\langle \mbox{tr}\, {\cal T}(n) \rangle$ in the
absence of time-reversal symmetry. Following the rules of the
semiclassical formalism, the average $\langle \mbox{tr}\,
{\cal T}(n) \rangle$ is determined by trajectory pairs
$\alpha$, $\beta$ that are piecewise paired throughout the entire
array of cavities. The trajectories can have small-angle self
encounters, at which the pairing between $\alpha$ and $\beta$ can be
changed. Each pair of
trajectories is classified by the number $m$ of times the trajectories
enter the $n$th cavity from the $(n-1)$st cavity. The $m$ segments 
of each trajectory in the $n$th cavity are labeled
$\alpha_1,\ldots,\alpha_m$ and $\beta_1,\ldots,\beta_m$.

As in Sec.\ \ref{sec:b2}, 
it will be sufficient to consider pairs of trajectories
in which each segment $\alpha_j$ is paired with the corresponding
segment $\beta_j$, $j=1,\ldots,m$. Trajectory pairs with self encounters 
in the $n$th cavity or with non-diagonal pairings in the $n$th cavity
give contributions to $\delta \langle \mbox{tr}\, {\cal T} \rangle$ 
that are a factor $1/g_{\rm c}$ smaller than the leading
contribution. For the diagonal pairing, the remaining $m$ segments 
of the trajectories $\alpha$ and $\beta$ that reside in the first
$n-1$ cavities precisely generate $\langle \mbox{tr}\,
{\cal T}(n-1) [\mbox{tr}\, {\cal R}(n-1)]^{m-1} \rangle$, where the
averaging brackets refer to variations of the Fermi energy for the
entire array of cavities. Hence, we find
\begin{eqnarray}
\langle \mbox{tr}\, {\cal T}(n) \rangle &=&
\sum_{m=1}^{\infty}
% \frac{1}{2^m g_{\rm c}^{m-1}}
 \frac{ \left\langle
  \mbox{tr}\,
  {\cal T}(n-1) [\mbox{tr}\, {\cal R}(n-1)]^{m-1} \right\rangle}{2^m g_{\rm c}^{m-1}}
\nonumber \\ &=&
\left\langle \frac{g_{\rm c} \mbox{tr}\, {\cal T}(n-1)}
{2 g_{\rm c} - \mbox{tr}\, {\cal R}(n-1)} \right\rangle.
\label{eq:Tavgdiff}
\end{eqnarray}
A schematic representation of trajectory pairs contributing to Eq.\
(\ref{eq:Tavgdiff}) with up to two self-encounters in the first $n-1$
cavities are shown in Fig.\ \ref{fig:avg}.
Using unitarity to express ${\cal R}$ in terms of ${\cal T}$ and
subtracting $\langle \mbox{tr}\, {\cal T}(n-1) \rangle$, one then 
finds
\begin{eqnarray}
\delta \langle \mbox{tr}\, {\cal T}(n) \rangle &=&
\langle \mbox{tr}\, {\cal T}(n) \rangle -
\langle \mbox{tr}\, {\cal T}(n-1) \rangle \nonumber \\ &=&
\frac{1}{g_{\rm c}} \langle (\mbox{tr}\, {\cal T}(n-1))^2 \rangle.
\end{eqnarray}
This is the same equation as one obtains from taking the average of
the increment $\mbox{tr}\, \delta {\cal T}$ in the stochastic approach.

\section{Field theory formulation}
\label{sec:field_theory}

In this section, we will approach the localization phenomenon from a
different perspective. We will use that the quantum dot array depicted
in Fig.~\ref{fig:1} supports hierarchies of different types of
field-fluctuations in a field-theoretic
description. These fluctuations reflect the fate of density
distributions in classical phase space under the dynamical evolution
of the system. 
Each of these fluctuations, thus, comes with a characteristic
`relaxation time', {\em i.e.,} a time scale on which the fluctuation
decays. (For example, fluctuations inhomogeneous in the sector of
phase space describing an individual quantum dot will decay on a time
scale comparable to the time of flight through the dot, etc.) In the
description of low energy phenomena such as the zero frequency (DC)
conductance, modes operating at short time scales can be treated
perturbatively. Their feedback into the sector of long time scales
then stabilizes a `low energy theory'. In the following, we will
derive a theory that is minimal in that it contains information
equivalent to that stored in the Fokker-Planck equation of
localization. The strategy pursued here parallels one applied
previously~\cite{kn:altland1996b} to the problem of dynamical localization in
the quantum kicked rotor (also known as the ``standard map''.) 
One difference is
that the spectrum of different modes encountered in the present
problem happens to be more complex.  Our logics also resembles that of
Ref.\ \onlinecite{mueller07}, where it had been shown that the relevant low
energy theory of an {\it ergodic} quantum system contains the
information otherwise stored in random matrix theory.

Technically, our discussion will be based on a formulation of the
array in terms of the ballistic nonlinear
$\sigma$-model.\cite{kn:muzykantski1995,A3S:NPh} A quadratic
approximation in energetically high lying modes generates an effective
low energy theory wherein each quantum dot is treated as a
structureless (`ergodic') entity. This theory will be equivalent to
the celebrated nonlinear $\sigma$-model of disordered quantum
wires,\cite{kn:efetov1997} a model that predicts Anderson
localization on large length scales. We will see that the parameters
stabilizing the hierarchical mode integration are the same as those
utilized in previous sections of this paper.

Conceptually, the hierarchical scheme is an alternative to an
indiscriminate perturbative integration over all modes in one
go. That latter scheme would be essentially equivalent (see
Ref.~\cite{mueller07} for a discussion in the context of spectral
statistics.) to a semiclassical expansion in terms of paired
trajectories. In this sense, the hierarchical mode integration
processes the information stored in the statistics of trajectories by
different means.

\subsection{Field theory of the quantum dot array}
\label{sec:field-theory-quantum}

Our starting point will be the description of the quantum dot array
in terms of the supersymmetric ballistic nonlinear
$\sigma$-model. This theory is obtained by averaging exact functional
representations of Green functions over an energy interval of width
$\Delta E$ centered around the uniform Fermi energy $E_F$ of the
array.  A subsequent saddle point approximation (stabilized in the
parameter $E_F/\Delta E \gg 1$) then obtains a field theory in
classical phase defined by
%\begin{widetext}
\begin{eqnarray}
\label{eq:5}
 Z &\equiv& \int DT\,\exp(-S[T]),
\end{eqnarray}
with
\begin{eqnarray}
 S[T] &=&  \frac{ \beta\pi \hbar \nu}{2} \int_\Gamma (d\bx)  \,{\rm tr}\Big(T\ast \Lambda \{
H,T^{-1}\}\Big) \nonumber \\ && \mbox{} + S_{\rm reg}[T].
\end{eqnarray}
%\end{widetext}
The integration variables in (\ref{eq:5}), $T(\bx) =
\{T^{\alpha\alpha'}(\bx)\}$ are (super) matrix valued fields defined
on shells $\Gamma=\{\bx|H(\bx)=E_0\}$ of constant energy in classical
phase space. Further, $\bx\equiv (\bq,\bp)$ where $\bq$ and $\bp$ are
coordinates and momenta, respectively, $H(\bx)$ is the Hamiltonian
function of the system, the integral over the energy-shell is
normalized to the (spatial) volume of the system, $\int_\Gamma
(d\bx)={\rm Vol}$, and $\nu$ is the single particle density of states
per volume, $\nu=1/(\Delta {\rm \,Vol})$. For time reversal and spin
rotation invariant systems (orthogonal symmetry class, $\beta=1$), the
`internal' structure of the matrices $T^{\alpha\alpha'}$ is described
by a composite index $\alpha=(a,r,t)$, where $a=+/-$ discriminates
between the advanced and retarded sector of the theory, $r={\rm
b},{\rm f}$ discriminates between commuting and anticommuting
sectors, and $t=1,2$ accounts for the operation of time reversal. Time
reversal symmetry reflects in the relation $(\tau
T^T\tau^{-1})(\bq,-\bp) = T^{-1}(\bq,\bp)$, where $\tau$ is a fixed
matrix whose detailed structure will not be of concern throughout. For
time reversal non-invariant systems (unitary symmetry, $\beta=2$) no
time-reversal structure exists and $\alpha=(a,r)$. In either case, the
matrices $T$ carry a coset space structure in the sense that
configurations $T$ and $TK$ are to be identified if $[K,\Lambda]=0$,
where $\Lambda=\sigma_3^{\rm ar}$ and ``ar'' stands for action in
advanced/retarded space. Finally, the regulatory action 
$$
 S_{\rm reg}[T]\equiv \delta \int_\Gamma (d\bx)\, 
 {\rm str}\Big( \Lambda T^{-1} \ast \Lambda T\Big),\ \
 \delta \downarrow 0
$$
determines the
`causality' of field fluctuations, but will otherwise not be of much
relevance throughout.

The fluctuation behavior of the fields $T$ in (\ref{eq:5}) is governed
by the classical Liouville operator $\{H,\;\}$ (where $\{\;,\;\}$ is
the Poisson bracket.) Quantum mechanics enters the problem through the
presence of Moyal products ``$\ast$'' in (\ref{eq:5}). In essence, this
product operation\cite{foot5} limits the maximum resolution of
the theory in phase space to scales of the order of a Planck cell.
% Specifically, the generator of classical time evolution $\{H,\;\}$
% acts on field configurations coarse grained in phase space over
% cells  $\sim \hbar^f$, rather than on mathematical points. 
In the following sections we reduce the above `bare' theory to
an effective low energy theory describing localization phenomena.

\subsection{Hierarchical mode integration}
\label{sec:hier-mode-integr}

\begin{figure}
\centerline{\resizebox{\hsize}{!}{\includegraphics{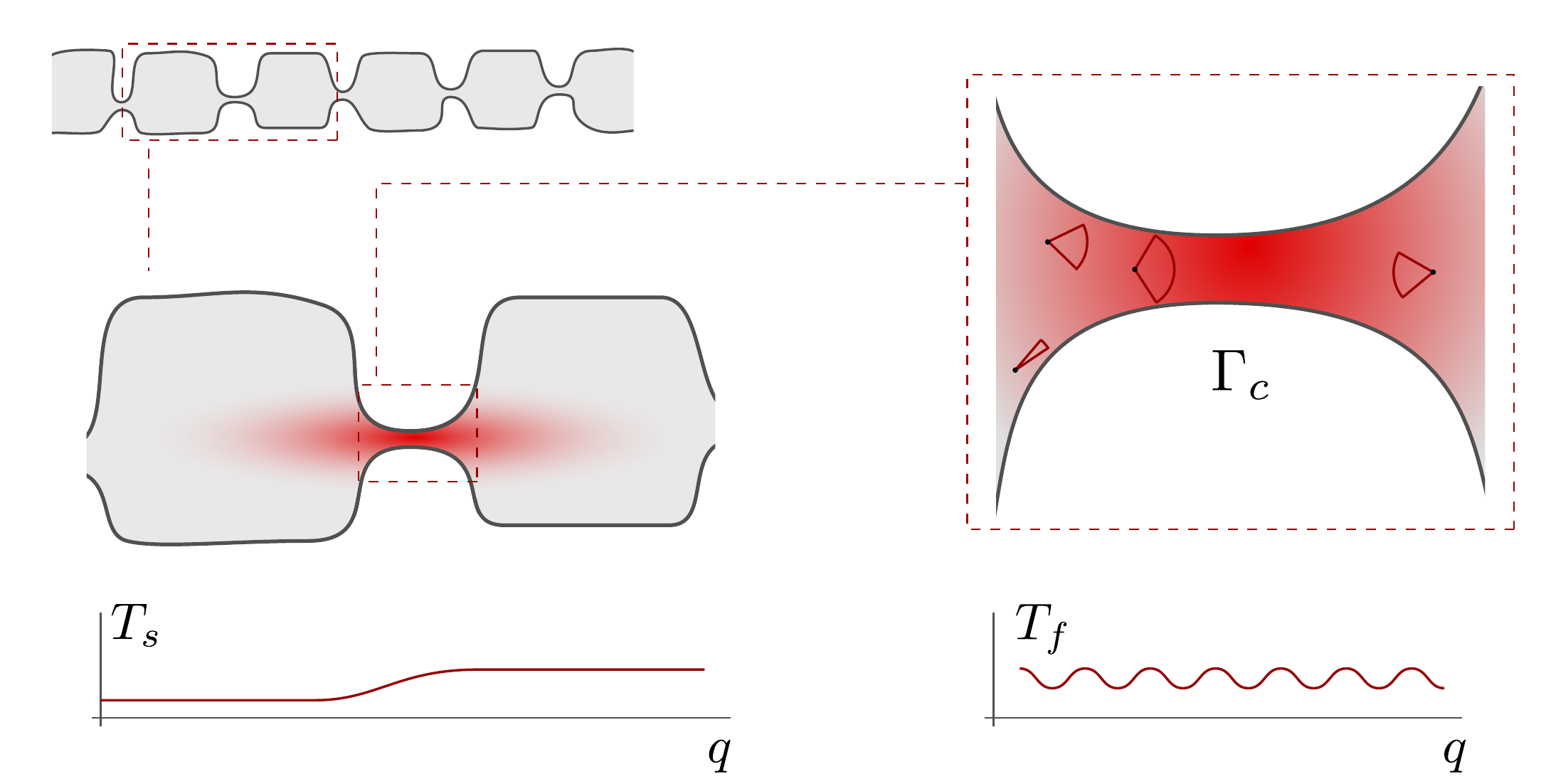}}}
\caption{(Color online) Top left: cartoon of an array
of weakly coupled quantum dots. Center left: cut region containing
two dots. The shaded region indicates the real space support of the
coupling phase space $\Gamma_c$.  Bottom left: plot indicating the
profile of weakly fluctuating field configurations: constancy
throughout each dot, gradual variation in the coupling
region. Center right: zoom into the coupling region. The phase space
region $\Gamma_c$ contains points $\bx=(\bq,\bn)$ that will migrate
between the two dots in short time, {\em i.e.}, without undergoing back
scattering. (The opening angles represent those directions $\bn$
that meet this criterion.) Bottom right: schematic plot of real
space profile of phase space fluctuations generating the inter-dot
coupling.}
\label{coupling_phase_space}
\end{figure}

Before turning to the technicalities of the mode integration, let us
describe the relevant hierarchies in qualitative terms: fluctuations
inhomogeneous in the phase space sector representing individual dots
are expected to relax on short time scales comparable to the time of
flight across the dot, $t_f$.\cite{foot6}
On the other hand,
relative fluctuations in the configuration of different dots can
survive up to time scales of the order of the dwell time $\tau_{\rm
D}\gg t_f$. To describe this hierarchical decay profile within our
field theory framework, we focus on a section of the array containing
two neighboring quantum dots
(cf. Fig.~\ref{coupling_phase_space}). The fields $T(\bx)$
representing phase space fluctuations in this subsystem may be
decomposed as $T=T_s T_f$, where $T_{s,f}$ are `slow' and `fast'
fluctuations, respectively. The slow fluctuations are (i) homogeneous
within each dot separately. In particular, (ii) they do not vary in
any spatial cross section transverse to the array, and (iii) do not
depend on momentum. However, (iv) the weakness of the inter-dot
coupling ($\tau_{\rm D} \gg t_f$) leaves room for gradual fluctuations
of the slow modes as we pass from one dot into the other. This
suggests a parameterization $T_s(\bx) = T_s(q)$, where $q$ is the
component of $\bq$ parallel to the longitudinal direction of the
two-dot system. Point (i) above means that $T_s(q\ll 0)=T_L$ and
$T_s(q\gg 0)=T_R$, where $T_{L,R}$ are the constant slow mode
configurations of the left and the right dot, respectively. As we are
going to check in a self consistent manner, (v) relative fluctuations
between left and right dot are suppressed (in a parameter of the order
of the number of transverse quantum channels supported by the
connector region) so that a leading order expansion in relative
fluctuations $T^{\vphantom{-1}}_L T_R^{-1}$ is
sufficient. (Conceptually, this expansion is equivalent to the
Kramers-Moyal expansion
employed in the derivation of the Fokker-Planck equation
above.)

The field $T_f$ encapsulates all other fluctuations, {\em i.e.}, fluctuations
that do not meet the criteria (i-v).  Generic fluctuations of this
type -- think of fast fluctuations deep inside the phase space of one
of the dots -- are strongly gapped and do not couple to the slow
fluctuations.  (Formally, this decoupling manifests itself in an
effective `orthogonality' between the modes representing these
fluctuations.) However, there is one sector of phase space,
$\Gamma_c$, in which fast and slow fluctuations talk to each
other. The domain $\Gamma_c$ includes all points in phase space which
pass from one dot to the other in a time of order $t_f\ll \tau_{\rm
D}$, much smaller than the typical dwell time (cf. the shaded area
in Fig.~\ref{coupling_phase_space}). This region is special in that it
overlaps with the domain of gradual variation of the slow fields
(cf. point (iv) above.) As we will see, a perturbative integration
over fast fields in $\Gamma_c$ effectively determines the slow field
coupling between the two dots.

To prepare the integration over the fast fields in $\Gamma_c$, we need
to bring the notation to a more explicit level: assuming that the bulk
dynamics is ballistic, $\{H,\;\} = v_F {\bf n}\cdot \nabla$ where
${\bf n}$ is the unit vector in momentum space, and $v_F$ the Fermi
velocity. Current conservation in the specular reflection at the system
boundaries translates to the effective boundary condition $f(\bx,\bn)
= f(\bx,\bar \bn)$, where $f$ is a phase function subject to the
action of $\{H,\;\}$, $\bx$ a boundary point and $\bar \bn$ the
direction vector with flipped normal component.

We next parameterize fluctuations as $T_{s,f}=\exp(W_{s,f})$, where
the field generators carry a block structure (in advanced/retarded
space), $$ 
W_{s,f}=\left(
\begin{smallmatrix}
 & B_{s,f}\\
  \bar B_{s,f}&
\end{smallmatrix}\right).
$$
(For further technical details on this
representation we refer to Ref.\ \onlinecite{kn:efetov1997}.) We now
substitute these generators into the action, expand to leading order
in $W_f$, and integrate. This leads to the effective action
\begin{equation}
\label{Seff}
S_{\rm eff}[T_s]= \frac{1}{2}\left \langle
\left(S^{(1)}[T_s,W_f]\right)^2\right\rangle + S_{\rm reg}[T_s],
\end{equation}
where $\langle \dots \rangle \equiv \int
DW_f\,\exp(-S^{(2)}[\openone,W_f])(\dots)$, and $S^{(n) }[T_s,W_f]$ is
of $n$th order in $W_f$. Due to the isotropy of $T_s$ in momentum
space and the linearity of the Poisson bracket in $\bn$, the action
does not contain a contribution of zeroth oder in the fast fields. The
dominant contribution to the coupling between fast and slow
fluctuations is given by the linear term,
\begin{align}
S^{(1)}[T_s,W_f]&= \beta \pi \hbar \nu v_F \int_{\Gamma_c} d\bq
d\bn\,n_\parallel \,{\rm str}\left[W_s(\bq,\bn) \Phi(q)\right]
\label{TsW}
\end{align}
where we have introduced the abbreviation $\Phi \equiv (\partial_q T_s)
\ast\Lambda T_s^{-1}$, the integration over the direction of momentum
is normalized as $\int d\bn =1$, $n_\parallel$ is the component of
$\bn$ parallel to the longitudinal direction and the Moyal product has
been omitted (which is permissible due to the general relation $\int
(d\bx) (f\ast g)(\bx) = \int (d\bx) (f\,g)(\bx)$ for the integrated
product of \textit{two} functions -- presently, matrix elements of
$\Phi$ and $W$ -- in phase space.)

Neglecting contributions of $\mathcal{O}(\Phi W_f^2)$ (as compared to
the $\mathcal{O}(\Phi W_f)$-terms above) the quadratic $W_s$-action is
given by
\begin{widetext}
\begin{eqnarray*}
S^{(2)}[T_s,W_f] &\simeq & S^{(2)}[\openone,W_f]
 \nonumber \\ &=& \beta \pi \hbar \nu 
\int_{\Gamma_c} d\bq d\bn \,{\rm str}(\bar B_f(\bq,\bn) (v_F \bn\cdot
\nabla + \delta ) B_f(\bq,\bn)).
\end{eqnarray*}
Fast field fluctuations can now be integrated out according to the
prescription $ \langle {\rm str}(A(\bq,\bn) B_f(\bq,\bn))\; {\rm
str}(\bar A(\bq',\bn') \bar B_f(\bq',\bn'))\rangle =
\delta(\bn-\bn')\Pi(\bq,\bq';\bn) $, where
$$
\Pi(\bq,\bq';\bn)\equiv \frac{1}{ \beta \pi \hbar \nu}  (v_F \bn\cdot
\nabla + \delta )^{-1}(\bq,\bq').
$$
Specifically, the effective action is given by
\begin{eqnarray*}
S_{\rm eff}[T_s] &=& ( \beta \pi \hbar \nu v_F)^2\int_{\Gamma_c} d\bq
d\bq'd\bn\, n_\parallel^2
%  \nonumber \\ && \mbox{} \times
\mbox{str}\, \left(\Phi_{21}(q) \Phi_{12}(q) \right) \Pi(\bq,\bq';\bn) 
\end{eqnarray*}
So far, we have not made reference to the specific properties of the
phase space region $\Gamma_c$. We now assume that the corridor
connecting the dots has wave-guide properties, in that it (a) does not
contain significant backscattering, and (b) the restriction of the
Liouville operator to $\Gamma_c$ has plane-wave like  eigenfunctions
characterized by a conserved  longitudinal/transverse momentum
$\bk_\perp/k$. We also assume that (c) the slow fields smoothly
interpolate between $T_L$ and $T_R$ in a region centered around the
longitudinal coordinate $q=0$. The presumed proximity $T_L \simeq T_R$
implies that $\Phi(q) \simeq \partial_q W(q)\Lambda$ can be linearized, where we
suppressed the slow field index `$s$' for notational
transparency. Under these circumstances, and noting that the
integration over $\bq_\perp$ implies a projection onto the
zero-momentum sector $\bk_\perp=0$, we obtain
%\begin{widetext}
\begin{align*}
S_{\rm eff}[T_s] &= -\beta \pi \hbar \nu S_c v^2_F\int_{-c}^c dq dq'
{\rm \,str}\left(\Phi_{21}(q)\Phi_{12}(q') \right)  \int
d\bn\, n^2_\parallel \int \frac {dk}{ 2\pi} \frac{e^{i k (q-q')}}{i v_F n_\parallel k+
  \delta} =\\
&= -\beta \pi \hbar \nu S_c v_F\int_{-c}^c dq dq'
{\rm \,str}\left(\Phi_{21}(q) \Phi_{12}(q') \right)  \underbrace{\int
d\bn\, |n_\parallel|\Theta\left( \frac{q-q'}{
    n_\parallel}\right)}_{={\rm const.}}=\\
&={\rm const.}\times \hbar \nu S_c v_F\, {\rm str}((\bar B_R-\bar
B_L)(B_R - B_L)),
\end{align*}
\end{widetext}
where the indices refer to ``ar''-space, ${\rm const.}=\mathcal{O}(1)$ is a constant, $S_c$ the
transverse cross section of the contact, and $B_{R,L}$ are the
generators of the slow fields in the left and the right dot,
respectively. Noting that the density of states per volume, $\nu \sim
m k_F^{d-2}$ (where $k_F = v_F m$ is the Fermi momentum) and $S_c
k_f^{d-1}\sim N$ is proportional to the number of transverse channels,
$N$, supported by the connector region, the
prefactor can be written as ${\rm const.}\times N \gg 1$, a number
which we assume large lest a semiclassical description of the contact
becomes meaningless. We also note that the quadratic form may be
replaced by its unique rotationally invariant generalization to the full field
manifold, ${\rm str}((\bar B_R-\bar
B_L)(B_R - B_L))\to \frac{1}{ 4}{\rm str}(Q_L Q_R)$, where $Q=T \Lambda
T^{-1}$. While a quadratic expansion of the latter expression reproduces the
bilinear term, the largeness of $N$ implies that typical values
contributing to the $B$-integration, $B\sim N^{-1/2}\ll 1$, which means
that non-linear contributions become inessential in the limit of large
channel numbers. We thus conclude that the coupling term can be
rewritten as
$$
S_{\rm eff}[Q]= {\rm const.} \times N {\,\rm str}(Q_L Q_R).
$$
Finally, the obvious  generalization of the above two-dot construction
to an entire quantum dot array reads as
\begin{equation}
\label{Qdotarray}
S_{\rm eff}[Q]= {\rm const.} \times \frac{1}{\tau_{\rm D} \Delta} \sum_{m} {\rm str}(Q_m Q_{m+1}),
\end{equation}
where $Q_m$ is the $Q$-matrix representing the $m$-th dot and we used
that the channel number $N\sim (\tau_{\rm D} \Delta)^{-1}$. Before
turning to the discussion of localization properties, a few remarks on
the construction above are in order. 
\begin{itemize}
\item Conceptually, the above reduction programs involves three steps:
1) identification of `low energy modes', {\em i.e.}, modes that decay on
the largest time scales of the problem, 2) identification of `high
energy', or quickly decaying modes, and 3) perturbative integration
over those fast modes that will conceivably couple to the slow
modes. In principle, that integration can be explicated for any
fluctuation in the problem. In practice, however, only few modes
will effectively couple to the slow degrees of freedom, and these
relevant fluctuations are best identified by semiclassical
considerations:
\item Semiclassically speaking, a `mode' represents the coherent
propagation of a retarded and an advanced Feynman amplitude along
classical trajectories in phase space. Locally, the semiclassical
dynamics of such composites is described by the Liouville operator,
as is manifest in the action (\ref{eq:5}). This trajectory
interpretation helps in identifying the relevant fast modes. E.g.,
in the system depicted in Fig.~\ref{coupling_phase_space}, the
coupling between the ergodic slow modes of each dot, $Q_m$, is
dominated by trajectories swiftly propagating from one dot to the
other, {\em i.e.}, modes emanating at phase space points $\bx\in \Gamma_c$.
\item Although this identification of fast modes rests on specific
model assumptions (no backscattering in the contact region, etc.),
the result (\ref{Qdotarray}) is reasonably universal. For example, a
somewhat more elaborate construction will show that the same action
describes connector regions containing chaotic scattering and momentum
relaxation by themselves. (The latter complication would manifest
itself in an altered value of $\tau_{\rm D}$, though.) Generally
speaking, the low energy physics of the system will reduce to
$S_{\rm eff}$, as long as the dots are isolated from each other in
the sense $t_f \ll \tau_{\rm D}$.
\end{itemize}

\subsection{Localization from the effective action (\ref{Qdotarray}) }
\label{sec:field-theor-fokk}
The effective action (\ref{Qdotarray}) is equivalent to a `lattice version' of the
diffusive nonlinear $\sigma$-model of quasi-one dimensional disordered
wires. 
Indeed, we may pass to a continuum limit
$$
 \frac{1}{\tau_{\rm D} \Delta} \sum_{m} {\rm
str}(Q_m Q_{m+1}) \to \frac{a}{\tau_{\rm D}
\Delta} \int_0^L dx \, {\rm str}(\partial Q \partial Q),
$$
where $Q_m \to Q(x)$ is replaced by a smooth field, $x=ma$, and $a$
the  spacing between the dots. Comparing to the
standard form of the diffusive model,\cite{kn:efetov1997} 
where the action is $\sim \xi \int dx\, {\rm
str}(\partial Q \partial Q)$ with $\xi$ the localization
length, we are led to the identification $\frac{a}{\tau_{\rm D}
\Delta}\sim \xi$. 
%\texttt{remarks on field theoretical fokker-planck equation?}

\section{Conclusion}
\label{sec:concl}

In the preceding sections we showed how a theory of Anderson
localization can be constructed for a sample in which the microscopic
electron dynamics is ballistic, rather than
disordered-diffractive. Our theory of ``ballistic Anderson
localization'' paraphrases the scaling approach to localization 
in disordered quantum wires of Dorokhov, Mello, Pereyra, and 
Kumar,\cite{kn:dorokhov1982,kn:mello1988} using the language of 
the trajectory-based semiclassical formalism. 
As noted in the introduction, the interest of constructing such a
semiclassical theory is not the structure of the theory itself or the
phenomena it explains. Like
most semiclassical theories of quantum corrections in the perturbative
regime, the structure of
the theory closely resembles the structure of its fully quantum
mechanical counterpart for disordered metals, whereas the observed
phenomena are the same in the ballistic and disordered cases. 
Instead, the interest of the
semiclassical theory is that it shows how quantum effects that were
known from disordered metals arise if the electron dynamics is 
ballistic. 

On hindsight it should not come as a surprise that a theory of
ballistic Anderson localization can be constructed by adapting the
derivation of the Dorokhov-Mello-Pereyra-Kumar equation for a
disordered quantum wire. This point is best made by 
reconsidering Dorokhov's original derivation.\cite{kn:dorokhov1982}
% This
%calculation establishes a stochastic connection between the scattering
%matrix product ${\cal T}^{\rm q} = S_{12}^{\rm q} S_{12}^{{\rm
%    q}\dagger}$ of disordered wires of
%lengths $L$ and $L + \delta L$ if $\delta L \ll l$, where $l$ is the
%elastic mean free path. 
In this derivation, impurity scattering is treated in the Born 
approximation. 
%One may be tempted to expect that there can be no
%classical analogue of this derivation, because the Born
%approximation has no 
%counterpart in classical mechanics. On the other hand, 
All quantum mechanical amplitudes are squared into 
quantum probabilities. Hence, it is sufficient if one 
can replace the quantum mechanical probabilities by classical
ones. 
%%
%%; ``Hikami-box''-like interference corrections do
%%not appear at any point in the calculation. Hence, for a
%%rederivation of
%%the DMPK equation using classical trajectories it is sufficient if one 
%%can replace the quantum mechanical probabilities by classical
%%ones. 
Such a replacement is a standard procedure when connecting quantum
mechanical and semiclassical theories. Its implementation for the
array of chaotic cavities is what is done here.
%It is also what makes this 
%article's theory of ballistic Anderson localization possible.

%%; it is nothing but the
%%``diagonal approximation'' in semiclassical language, which is
%%used, {\em e.g.},
%%when replacing the ``Cooperon'' and ``diffuson'' propagators in a
%%quantum-mechanical description of perturbative quantum interference
%%corrections by their classical counterparts
%%\cite{kn:jalabert1990,kn:baranger1993,kn:baranger1993b,kn:aleiner1996}. 
%%Connecting quantum phenomena in disordered and
%%ballistic conductors is complicated only if Hikami-box-like features
%%appear in the quantum theory \cite{kn:aleiner1996}, 
%%but these are absent in Dorokhov's derivation.
%%
%For the standard geometry in which the DMPK equation is derived, a
%disordered wire to which a slice of thickness $\delta L \ll l$ is
%added, $l$ being the mean free path, one does not arrive at a suitable
%semiclassical description if one simply replaces quantum probabilities
%by classical ones. The problem is that the condition $\delta L \ll l$
%can not be reconciled with the condition that the electron dynamics be
%chaotic in the added slice. We here circumvented this problem by
%passing to an array of chaotic cavities and adding cavities one by
%one. This has the advantage that the electron dynamics in each added
%cavity is chaotic and ergodic, but the classical probability
%of reflection from a cavity is no longer small. The latter 
%complication can be overcome, however, if one includes multiple
%returns and up to one small-angle encounter in the
%added cavity.

It is interesting to observe that, while small-angle encounters form a
crucial link in our understanding of quantum interference corrections
in ballistic conductors, their role is very limited in our description
of localization in quasi-one dimension: They serve to cancel spurious
contributions from trajectories that enter the last cavity of the
array more than twice. (In Dorokhov's original approach such processes
are excluded automatically because of the condition that the length of
the wire is increased by an amount $\delta L$ much smaller than the
mean free path $l$.\cite{kn:dorokhov1982,kn:mello1988})
%This situation is not special for the semiclassical theory:
%As we discussed above, it is also a feature of Dorokhov's fully 
%quantum mechanical theory of 
%localization. Upon adding the $n$th cavity to the
%array of cavities one simply increases the classical return
%probability, thus enhancing the interference corrections to
%transport due to small-angle encounters in the first 
%$n-1$ cavities. Small-angle encounters in
%the $n$th cavity are not needed for a theory of localization.
Implicitly, small-angle encounters do play a much more important role in our
theory, however, because they help to preserve unitarity in the
semiclassical theory. Unitarity is a key ingredient of both the
quantum-mechanical derivation of the DMPK equation and the present
semiclassical derivation. We note that unitarity
has played an important role in other extensions of the
semiclassical framework beyond its previously
assumed domain of validity: it is
used to relate weak localization and enhanced backscattering, thus
enabling a semiclassical description of weak localization without
reference to small-angle
encounters,\cite{kn:baranger1993,kn:baranger1993b,kn:argaman1995}
and it is used
to obtain an alternative expression for the spectral form factor, 
allowing its calculation in the
non-perturbative regime by considering periodic orbits of duration
below the Heisenberg time only.\cite{kn:keating2007}

We have also shown that the dynamical information entering the
semiclassical approach can be processed by different means to derive
an effective low energy field theory of the system. This latter
approach is based on the concept of `modes', fluctuations in phase
space decaying on parametrically different time scales. A successive
integration over short lived modes stabilizes an effective action of
the most persistent modes in the fluctuation spectrum. In the present
context, that low energy limit turned out to be equivalent to the
diffusive nonlinear $\sigma$-model of disordered wires. While the
field theory approach is arguably less explicit than the direct
classification of trajectories, it enjoys the advantage of high
computational efficiency, a paradigm previously exemplified on the
problem of universal spectral correlations.\cite{mueller07} For
example, the above mentioned condition of unitarity, as well as the
symmetries relating trajectories to their time reversed are built into
the field theory approach from the outset; there is no need for
explicit bookkeeping in terms of encounter processes. The price to be
payed for this compactness in the description is a higher level of
abstraction, though.

%In the perturbative regime, trajectory-based semiclassical theories 
%of quantum transport in ballistic conductors often mirror theories of 
%quantum transport in disordered conductors. An example is the
%semiclassical theory of weak localization in ballistic chaotic
%cavities \cite{kn:aleiner1996,kn:richter2002}, which can be seen as a
%``translation'' of the diagrammatic perturbation theory of the
%same phenomenon in disordered quantum dots. In this translation, the
%role of the ``Cooperon propagator'' in the diagrammatic perturbation
%theory is taken by a pair of time-reversed classical
%trajectories, while the ``Hikami box'' is mapped to a small-angle
%encounter of classical trajectories.
%%In spite of the similarity of the structure of the two 
%%theories, it took more than a decade until a complete and reliable
%The situation is no different for the semiclassical
%theory of Anderson localization that is reported here: It, too, can be
%considered a ``translation'' of a theory of Anderson localization in
%disordered conductors. In our case, this theory is the scaling theory
%pioneered by Dorokhov \cite{kn:dorokhov1982} and Mello, Pereyra, and
%Kumar \cite{kn:mello1988}. In view of the similarity between
%theories of ballistic and disordered conductors in the perturbative
%regime this does not come as a surprise. Indeed, the main message of this
%communication is not about the structure of the semiclassical theory, 
%but about the fact such a theory is possible at all!

\section*{Acknowledgments}

We thank Fritz Haake for bringing this problem to our
attention. This
work was supported by the Packard Foundation, the Humboldt Foundation,
the NSF under grant no.\ DMR 0705476, and by the
Sonderforschungsbereich SFB/TR 12 of the Deutsche Forschungsgemeinschaft.

%\bibliography{refs}

\begin{thebibliography}{54}
\bibitem{foot4}
{This need not be
immediately obvious if time-reversal symmetry is present
($\beta=1$), since Eq.\ (\ref{eq:calTstat}) contains the matrix
${\cal G}^{\rm q}$, which can not be expressed in terms 
of ${\cal T}^{\rm q}$. Still, the stochastic process defined by 
Eqs.\ (\ref{eq:ttdiff}) fully describes the evolution of the
eigenvalues of ${\cal T}^{\rm q}$, which can be verified by
considering 
the mean change of $\mbox{tr}\, ({\cal T}^{\rm q})^{m}$ or of 
a product of such traces, which contains ${\cal G}^{\rm q}$ in the 
combination ${\cal G}^{\rm q} ({\cal T}^{*})^n {\cal
G}^{{\rm q}\dagger} = ({\cal T}^{\rm q})^{n+2} - ({\cal T}^{\rm
q})^{n+3}$ only, $n=0,1,2,\ldots$.}
\bibitem{foot5}
{The Moyal product between two phase
space functions $A$ and $B$ is defined as
% \begin{equation}
%   \label{eq:3}
%   (A\ast B)(\bx) = \int {d^f \bx_1\over (\pi \hbar)^f} {d^f \bx_2\over
%     (\pi \hbar)^f} \,e^{{2i\over \hbar} \bx_1^T I
%   \bx_2} A(\bx+\bx_1) B(\bx+\bx_2),
% \end{equation}
% where $I \equiv \left(
% \begin{smallmatrix}
%   0 &\openone\cr -\openone & 0
% \end{smallmatrix}
% \right)$ is the symplectic unit matrix.  (For all practical purposes,
%the definition (\ref{eq:3}) will be more convenient than the standard
%
$ (A\ast B)(\bx)=\exp(\frac{1}{2}{i\hbar}\partial^T_{\bx'} I
\partial^\vd_\bx)\; A(\bx')B(\bx)\big|_{x=x'}$, where $I \equiv 
% \left(
%  \begin{smallmatrix}
%    0 &\openone\cr -\openone & 0
%  \end{smallmatrix}
%  \right)
$ is the symplectic unit matrix.}
\bibitem{foot6}
{In fact, the phase space of
individual dots supports a small set of fluctuations that decay on
time scales longer than the generic $t_f$: the probability
$\Pi(\bx,\bar \bx',t)$ to propagate from a phase space point $\bx$
to the time reversed of a closeby point $\bx'$ relaxes on scales
$\sim \lambda^{-1} \ln(E_F t_f/us)$, where $\lambda$ is the dominant
Lyapunov exponent of the system and $s$ and $u$ are the locally most
stable and unstable coordinate of the point $\bx'$ in a coordinate
system that has $\bx$ as its center. Since the phase space
resolution of the quantum theory is limited by $us\sim \hbar$, these
long time memory effects decay on time scales of the order of the
Ehrenfest time $\tau_{\rm E} \sim \lambda^{-1} \ln(S/us)$. Thus, the
decoupled dots have relaxed into a fully ergodic configuration on
time scales $\tau_{\rm E}< \tau_{\rm D}$.}
\end{thebibliography}

\end{document}